\documentclass[
twocolumn,
english,
superscriptaddress,
amsmath,
amssymb,
floatfix,
longbibliography,
floats,
aps,
prx,
10]{revtex4-2}
\usepackage{graphicx}
\usepackage{dcolumn}
\usepackage{bm}
\usepackage{indentfirst}
\usepackage{graphicx} 
\usepackage{amsmath,amsfonts,graphicx,color,bbm,tikz,bm,setspace}
\usepackage{bbold}
\usepackage{float}
\usepackage[dvipsnames]{xcolor}
\begin{document}
\preprint{APS/123-QED}

\title{{Noninvertible Kramers-Wannier duality symmetries for the discrete-time quantum Ising chain}}
\author{Akash Sinha}
\email{akash26121999@gmail.com}
\affiliation{\it Department of Physics, School of Basic Sciences,\\ Indian Institute of Technology, Bhubaneswar, 752050, India}

\author{Pramod Padmanabhan}%
\email{pramod23phys@gmail.com,}
\affiliation{\it Department of Physics, School of Basic Sciences,\\ Indian Institute of Technology, Bhubaneswar, 752050, India}


\author{Vladimir Korepin}
\email{vladimir.korepin@stonybrook.edu}
\affiliation{C. N. Yang Institute for Theoretical Physics, \\ Stony Brook University, New York 11794, USA}



\begin{abstract}
\textit{Integrable trotterization} provides a method to evolve a continuous time integrable many-body system in discrete time, such that it retains its conserved quantities. Here we explicitly show that the first order trotterization of the critical {\it transverse field Ising model} is integrable. The discrete time conserved quantities are obtained from an inhomogeneous transfer matrix constructed using the {\it quantum inverse scattering method}. The inhomogeneity parameter determines the discrete time step. We then focus on the non-invertible {\it Kramers-Wannier} duality-symmetry for the trotterized evolution. We find that the discretization of both space and time leads to a doubling of these duality operators. They account for discrete translations in both space and time. As an interesting application, we find that these operators also provide maps between trotterizations of different orders. This helps us extend our results beyond the trotterization scheme and investigate the Kramers-Wannier duality-symmetry for finite time Floquet evolution of the critical transverse field Ising chain. {Finally, we investigate how these non-invertible operators shape the phase diagram of the discrete-time evolution. This question is particularly interesting in the Floquet setting, which is known to host a richer phase structure than its undriven counterpart. We systematically construct the necessary operators which relate different phases away from criticality for both trotterized and Floquet evolutions.}
\end{abstract}

\maketitle

\section{Introduction}
For close to a century now, quantum integrable systems (Heisenberg models, Ising models, Potts models, Hubbard models etc.) have played a central role in helping us understand many-body systems. Their extensive set of conserved charges strongly constrains dynamics, influencing both the equilibrium \cite{Rigol2007,Essler2016} and the non-equilibrium \cite{calabrese2016introduction,CastroAlvaredo2016,Bertini2016} properties. A key feature of the integrable systems is their exact solvability, using techniques like the coordinate Bethe ansatz \cite{bethe1931theorie} and the quantum inverse scattering method (QISM) \cite{faddeev1996algebraic,Korepin1993QuantumIS,Sklyanin1982QuantumVO,faddeev1980quantum}. In recent times this property has been exploited to benchmark the performance of quantum computers and testing quantum simulation algorithms \cite{maruyoshi2023conserved,farreras2025simulation,chowdhury2025quantum}. To carry out these simulations we need to decompose the continuous time evolution into a sequence of discrete unitaries. One way to do this is with the so-called \textit{trotterization}\cite{lloyd1996universal,lanyon2011universal,childs2019nearly,pastori2022characterization,PhysRevX.8.011044,choi2025quantum}, based on the Suzuki-Trotter formula \cite{trotter1959product,suzuki1976generalized,suzuki1985decomposition}
\begin{eqnarray}\label{eq:1stTrot}
    e^{-{\rm i}H_{}t}=\lim_{n\to\infty}\left(e^{-{\rm i}H_A t/n}e^{-{\rm i}H_B t/n}\right)^n,
\end{eqnarray}
where $H_{}=H_A+H_B$ is the Hamiltonian generating the time evolution of the many-body system. Usually, the total Hamiltonian is decomposed in such a way that each local term from a particular component of the Hamiltonian commutes with all the other local terms from the same component. 

Such naive discretization procedures often destroy the integrals of motion of the continuous-time evolution. This beats the purpose of exactly solvable models in the context of quantum simulations. Therefore, it is essential to figure out discretization methods that preserve the integrable structure. The QISM framework helps identify certain integrable models, which remain integrable even after trotterization, i.e. the corresponding quantum circuit also comes with an extensive number of conserved quantities. Such a scheme was first put forward in \cite{vanicat2018integrable,ljubotina2019ballistic} for the integrable Heisenberg spin chains. Following this, different aspects of such integrable trotterization were studied in \cite{medenjak2020rigorous,aleiner2021bethe,giudice2022temporal,claeys2022correlations,vernier2023integrable,vernier2024strong,paletta2025integrability,Calliari_2025}.  Notably, in \cite{maruyoshi2023conserved} the integrable trotterization of the $XXX$-Heisenberg spin chain has been implemented on real quantum hardware, and the decay of conserved charges under Trotter error and noise has been measured. These works show that integrable trotterization is a robust way to benchmark quantum simulators.
  
In this paper we study the integrable trotterization of the critical transverse-field Ising model (TFIM), especially from the point of view of symmetries of the trotterized evolution. In our recent work \cite{Sinha:2025wqf}, we proved that the TFIM can be obtained by QISM, thus rendering it Yang-Baxter integrable. However, this does not prove the integrability of the trotterized time evolution. Adopting the construction given in \cite{Sinha:2025wqf}, we first show that the periodic, critical TFIM admits an integrable trotterization. The resulting quantum circuit factorizes into local, unitary quantum gates and commutes with an appropriate transfer matrix. Among the other conserved quantities, we find the discrete-time generalizations of the well-known \textit{Kramers-Wannier}(KW) duality transformation. 

Originally introduced in \cite{KW1941-1,KW1941-2}, the KW duality played an instrumental role in estimating the critical temperature of the $2$-D classical Ising model by relating the `high' and `low' temperature expansions of the partition function. In the setting of the $1$-D TFIM spin-chain, which is weakly equivalent to the $2$-D classical Ising model, the KW duality provides a map between the ferromagnetic (ordered) and the paramagnetic (disordered) phases. Since these two phases have completely different ground-state structures, the mapping is non-invertible. Furthermore, at the critical point between the two phases, where the system becomes self-dual, the KW duality becomes a (non-invertible) symmetry \cite{aasen2016topological,aasen2020topological,Seiberg2024_NonInvertibleLSM,shao2023s,Zhang_2025,Sinha:2025wqf}. 

One way to implement this non-invertible duality is by introducing a dual Hilbert space \cite{aasen2016topological,aasen2020topological} in which the degrees of freedom live on the links between two adjacent sites of the original spin-chain. The KW duality operator then provides a map from the original Hilbert space to this dual Hilbert space. Notably, in \cite{10.21468/SciPostPhys.16.3.075}, the authors have extended this construction to the Floquet TFIM. We will not, however, follow this particular approach. Instead, we will be working with a different, but related framework, where the duality operators act only on the original Hilbert space of $N$ spin-1/2 degrees of freedom. This distinction is important since it dictates the algebra of the non-invertible duality operator. Crucially, when one works with a single Hilbert space, the algebra of the non-invertible duality operator mixes with the lattice translation \cite{seiberg2024majorana,Seiberg2024_NonInvertibleLSM,shao2023s}.

However, as we shall show explicitly, the usual, continuous-time KW duality no longer remains a symmetry of the discrete-time trotterized evolution of the critical TFIM. Interestingly, the integrable construction allows us to define appropriate KW duality operators, which perform the necessary duality transformations and also commute with the discrete time evolution. We expose the algebra of these discrete-time KW operators and show that it involves spatial as well as temporal translations.

Our findings are organized as follows. We begin with the integrable trotterization of the critical TFIM in  Section \ref{sec:IntTrottCTFIM}. The key step is to trotterize the time-evolution of a Majorana fermionic chain, as presented in Subsection \ref{sec:IntTrottMaj}. Subsequently, we employ the \textit{Jordan-Wigner} (JW) transformation to map the Majorana chain to the critical TFIM in subsection \ref{sec:MajJWTFIM}. The integrable quantum circuit for the critical TFIM is given in Equation \eqref{eq:FinalCircuit}. In Section \ref{sec:DisKW} we construct the KW duality-symmetry of the trotterized time-evolution of the critical TFIM. We find, in Subsection \ref{sec:DisKWDuSy}, appropriate generalizations of the KW duality-symmetry operator of the continuous-time critical TFIM. These operators, as given in \eqref{eq:DiscreteKWexp}, are non-invertible, perform the required duality transformations and commute with the trotterized time-evolution operator \eqref{eq:DualDisTFIM}. The algebra generated by these non-invertible symmetries is written down in Subsection \ref{sec:AlgDisKW}. In Section \ref{sec:KWFloquet}, we extend our discussion to the Floquet evolution of the critical TFIM. We show that, for a certain range of the Floquet time-period, the resulting circuit still remains Yang-Baxter integrable. In particular, we investigate the notion of appropriate KW duality for some specific Floquet time evolutions. {In the subsequent Section \ref{sec:phase-structure}, we discuss how the non-invertible duality-symmetries influence the phase-structure of the discrete-time evolutions. We examine both trotterized and Floquet cases in detail.} In Section \ref{sec:conclu}, we summarize our main findings and suggest future directions. This work is supplemented with three appendices. In Appendix \ref{app:transfercommutation}, we briefly discuss the procedure to obtain the commuting transfer matrices by appropriately performing the partial trace over the auxiliary Hilbert space. In the next Appendix \ref{app:LocCon} we give the expressions of the first few local conserved quantities that we obtain from the inhomogenous transfer matrix. Finally, in Appendix \ref{app:Onsager} we describe how the Onsager algebra can be obtained from QISM.

\section{Integrable trotterization of the critical TFIM}\label{sec:IntTrottCTFIM}
\begin{figure}[H]
    \centering
    \includegraphics[width=1\linewidth]{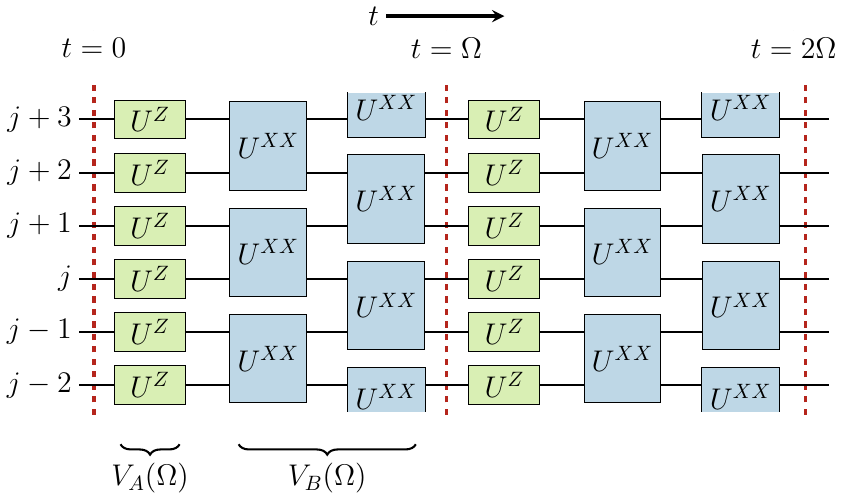}
    \caption{A schematic description of the discrete time-evolution \eqref{eq:TROTTAPPROX}, obtained by successively applying the elementary quantum circuit ${\sf V}(\Omega)=V_A(\Omega)V_B(\Omega)$. We dropped the argument $\Omega$ from the quantum gates $U^Z(\Omega),U^{XX}(\Omega)$ to avoid having a clumsy diagram.}
    \label{fig:circuitschematic}
\end{figure}

The critical TFIM is described by the Hamiltonian
\begin{eqnarray}\label{eq:TFIM ham}
    &&{\sf H}_{\rm TFIM}=H_A+H_B,\nonumber\\
    &&H_A=-\sum_{j=1}^NZ_j,\quad H_B=-\sum_{j=1}^NX_jX_{j+1},
\end{eqnarray}
with periodic boundary condition $N+1\equiv 1$. Here $X,Z$ are the Pauli matrices and ${\cal O}_j=\mathbb{1}_2^{\otimes j-1}\otimes{\cal O}\otimes\mathbb{1}_2^{\otimes N-j}$. An appropriate quantum circuit that can approximate the continuous-time evolution is given by
\begin{eqnarray}\label{eq:Gates}
    &&{\sf V}(\Omega)=V_A(\Omega)V_B(\Omega),\nonumber\\
    && V_A(\Omega)=\prod_{j=1}^NU^{\rm Z}_j(\Omega),~ V_B(\Omega)=\prod_{j=1}^NU^{\rm XX}_j(\Omega),
\end{eqnarray}
with the elementary quantum gates
\begin{eqnarray}\label{eq:approxGates}
    U^{\rm Z}_j(\Omega)=\frac{{\mathbb 1}+{\rm i}\Omega Z_j}{1+{\rm i}\Omega},\quad U^{\rm XX}_j(\Omega)=\frac{{\mathbb 1}+{\rm i}\Omega X_jX_{j+1}}{1+{\rm i}\Omega}.
\end{eqnarray}
Here $\Omega$ is the duration of the discrete Trotter step. Repeated application of the unitary ${\sf V}(\Omega)$ (Fig. \ref{fig:circuitschematic}) approximates the continuous-time evolution as
\begin{eqnarray}\label{eq:TROTTAPPROX}
    {\sf V}(\Omega)^n=e^{-{\rm i}t{\sf H}_{\rm TFIM}}+{\cal O}(t^2/n),\quad \Omega=t/n.
\end{eqnarray}
The above choice of the gates is motivated by the fact that for small enough $\Omega$, we essentially have
\begin{eqnarray}
    U^Z_j(\Omega)\simeq e^{{\rm i}\Omega Z_j},\quad U^{XX}_j(\Omega)\simeq e^{{\rm i}\Omega X_jX_{j+1}}.
\end{eqnarray}
Furthermore, since reducing the size of the time-step $\Omega$ for a fixed time-interval $t$ effectively lowers the Trotter error \eqref{eq:TROTTAPPROX}, the choice of the local gates in \eqref{eq:approxGates} indeed provides a valid approximation of $\exp[-{\rm i}t{\sf H}_{\rm TFIM}]\simeq{\sf V}(\Omega)^n$ for small $\Omega$ and large $n$, keeping $t=n\Omega$ stationary. The above circuit is free-fermionic \cite{jozsa2008matchgates} and certain conserved quantities for it can be obtained by exploiting the \textit{Onsager algebra}, as demonstrated in \cite{prosen1998new,Prosen2000ExactTF,gritsev2017integrable}. More generally, one can consider two different and not necessarily small time periods $\Omega_{A,B}$ for the two Hamiltonians $H_{A,B}$, respectively. The resulting binary Floquet drive
\begin{eqnarray}
    {\sf V}(\Omega_{A},\Omega_B)=e^{-{\rm i}\Omega_AH_A}e^{-{\rm i}\Omega_BH_B},
\end{eqnarray}
has been studied in detail in \cite{khemani2016phase,PhysRevB.93.245145,von2016phase,berdanier2018floquet}\footnote{To be precise, the authors in \cite{khemani2016phase,PhysRevB.93.245145,von2016phase,berdanier2018floquet} study the evolution ${\sf V}(\Omega_A,\Omega_B)$ without the boundary term $X_NX_1$.}. This system is shown to exhibit a rich phase structure. In fact, the case  $(\Omega_A=\Omega_B=\Omega)$ lies at the boundary between two different phases.
We wish to establish that the circuit ${\sf V}(\Omega)$ in \eqref{eq:approxGates} can be obtained by QISM, thus making it Yang-Baxter integrable. Our starting point is the non-local, Majorana fermionic $R$-operator
\begin{eqnarray}\label{eq:R-op}
    R_{a,b}(\lambda)=\left(\frac{\gamma_a-\gamma_b}{\sqrt{2}}\right)\left(\frac{{\mathbb 1}+\tanh(\lambda)\gamma_a\gamma_b}{1+{\rm i}\tanh(\lambda)}\right),
\end{eqnarray}
which solves the spectral parameter dependent Yang-Baxter equation (YBE)
\begin{eqnarray}
    &&R_{a,b}(\lambda-\mu)R_{a,c}(\lambda)R_{b,c}(\mu)=\nonumber\\
    &&\hspace{3cm}R_{b,c}(\mu)R_{a,c}(\lambda)R_{a,b}(\lambda-\mu).
\end{eqnarray}
Here $\gamma$'s are the Majorana fermions, satisfying $\{\gamma_a,\gamma_b\}=2\delta_{a,b}$. Notably, $R_{a,b}(0)=(\gamma_a-\gamma_b)/\sqrt{2}=:P^-_{a,b}$ is the non-local, Majorana fermionic permutation operator, introduced in \cite{Sinha:2025wqf}. Considering $R(\lambda)=P^-\check{R}(\lambda)$, we have $\check{R}(\lambda)^\dagger\check{R}(\lambda)=\mathbb{1}$. This allows the construction of local, unitary quantum gates using $\check{R}$-operator. One now can build the monodromy operator
\begin{eqnarray}\label{eq:inhomMono}
    T_{a(b)}(\lambda|\{\eta\})={{\cal C}_N}\prod_{j=2N}^1R_{a(b),j}(\lambda-\eta_{j}),
\end{eqnarray}
which satisfies the $RTT$-relation
\begin{eqnarray}\label{eq:RTT}
    &&R_{a,b}(\lambda-\mu)T_a(\lambda|\{\eta\})T_b(\mu|\{\eta\})=\nonumber\\
    &&\hspace{2.5cm}T_b(\mu|\{\eta\})T_a(\lambda|\{\eta\})R_{a,b}(\lambda-\mu).
\end{eqnarray}
{The prefactor ${\cal C}_N=\sqrt{2}e^{\frac{2\pi{\rm i}N}{8}}$ is introduced to achieve proper normalization.} Here $\{\gamma_a,\gamma_b\}$ and $\{\gamma_j|j=1,\cdots,2N\}$ are the auxiliary and the physical Majorana fermions, respectively. We use the notation $\{\eta\}$ to collectively describe the set of the inhomogeneities $\{\eta_1,\cdots,\eta_{2N}\}$. The above $RTT$-relation leads to the commuting set of transfer matrices $\tau(\lambda|\{\eta\})={({\rm dim}~{\cal H}_{ab})^{-1}}{\rm tr}_{{\cal H}_{ab}}\left[T_{a(b)}(\lambda|\{\eta\})\right]$ as
\begin{eqnarray}
    &&[\tau(\lambda|\{\eta\}),\tau(\mu|\{\eta\})]=0,
\end{eqnarray}
for different spectral parameters $\lambda,\mu$, with a fixed set of inhomogeneities $\{\eta\}$. Here the partial trace is over the auxiliary Hilbert space ${\cal H}_{ab}$. Since we are working with fermions, performing partial trace becomes somewhat involved. We give detailed steps to do so in Appendix \ref{app:transfercommutation}. The reader may refer to \cite{zhu2025systematic} for some interesting results, involving inhomogeneous transfer matrices constructed from the non-local, fermionic $R$-operators, similar to the one given in \eqref{eq:inhomMono}.

\subsection{Integrable trotterization of a Majorana chain}\label{sec:IntTrottMaj}
To begin with, we briefly recall the results for the completely homogeneous case $\eta_j=0,~\forall j$. The Hamiltonian can be obtained by taking the derivative of the logarithm of the completely homogeneous transfer matrix as
\begin{eqnarray}\label{eq:HamMajInt}
    &&H={\rm i}\frac{{\rm d}}{{\rm d}\lambda}\ln\tau(\lambda|0)\Big|_{\lambda=0}={\rm i}\sum_{j=1}^{2N}(-1)^{\delta_{j,2N}}\Gamma_j\Gamma_{j+1}.
\end{eqnarray}
Here $\Gamma_j$'s are the representations of the Majorana fermion $\gamma_j$'s on the physical Hilbert space and consequently satisfy the same algebra $\{\Gamma_j,\Gamma_k\}=2\delta_{j,k}{\mathbb 1},~\forall j,k=1,\cdots,2N$. The above Hamiltonian describes a system of $2N$ Majorana modes on a chain with anti-periodic boundary condition \cite{seiberg2024majorana,Sinha:2025wqf}. A conserved quantity of particular interest is the twisted translation operator ${\cal U}:={\tau(0|0)}$, which explicitly is given as
\begin{eqnarray}\label{eq:TwistTrans}
    &&{\cal U}={e^{-\frac{3\pi{\rm i}N}{4}}}\left(\frac{\Gamma_1-\Gamma_2}{\sqrt2}\right)\cdots\left(\frac{\Gamma_{2N-1}-\Gamma_{2N}}{\sqrt2}\right)\Gamma_{2N},\nonumber\\
    &&{\cal U}\Gamma_j{\cal U}^{-1}=\Gamma_{j+1},~{\cal U}\Gamma_{2N}{\cal U}^{-1}=-\Gamma_{1},~{\cal U}^{2N}\propto {\cal P},
\end{eqnarray}
with $j=1,\cdots,2N-1$ and ${\cal P}=\prod_{j=1}^{2N}\Gamma_j$ being the fermionic parity. The key step to go from the above fermionic Hamiltonian to the critical TFIM \eqref{eq:TFIM ham} is by exploiting the JW transformation to represent the Majorana fermions in terms of spin-1/2 operators. However, before doing that, we wish to investigate the trotterization of the fermionic Hamiltonian itself. Subsequently we can apply the JW transformation to obtain the trotterized evolution of the critical TFIM. 

To simulate the dynamics generated by the Hamiltonian in Equation \eqref{eq:HamMajInt}, we need to break the time evolution $\exp[-{\rm i}Ht]$ into a sequence of quantum gates. A possible way to achieve this is by introducing the quantum circuit
\begin{eqnarray}\label{eq:TrottCirc}
    &&{\cal V}(\Omega) =\prod_{j=1}^NU_{2j-1,2j}(\Omega) \prod_{j=1}^NU_{2j,2j+1}(\Omega),
\end{eqnarray}
with the local quantum gates 
\begin{eqnarray}
    U_{j,j+1}(\Omega) \simeq \frac{{\mathbb 1}+(-1)^{j,2N}\Omega  \Gamma_{j}\Gamma_{j+1}}{1+{\rm i}\Omega}.
\end{eqnarray}
The approximate time evolution becomes $${\cal V}(\Omega)^n=\exp[-{\rm i}Ht]+{\cal O}(t^2/n),\quad \Omega=t/n.$$ We shall show that the corresponding ${\cal V}(\Omega) $ can be obtained from a suitable transfer matrix, thus making it integrable. 

Let us now recall the inhomogeneous {monodromy operator} in \eqref{eq:inhomMono} and consider the simple non-trivial inhomogeneity, where all the even indices and all the odd indices are associated with the inhomogeneity parameter $\omega/2$ and $-\omega/2$, respectively. To be explicit, the monodromy operator takes the form
\begin{eqnarray}\label{eq:inhomMonoIsing}
    &&T_{a(b)}(\lambda|\omega)={{\cal C}_N}\prod_{j=2N}^1R_{a(b),j}\left(\lambda-(-1)^{j}\frac{\omega}{2}\right).
\end{eqnarray}
The transfer matrix is obtained as $\tau(\lambda|\omega)={({\rm dim}~{\cal H}_{ab})^{-1}}{\rm tr}_{{\cal H}_{ab}}\left[T_{a(b)}(\lambda|\omega)\right]$, which commute for two different spectral parameters $[\tau(\lambda|\omega),\tau(\mu|\omega)]=0$.
The expression for the transfer matrix becomes particularly simple when one considers two special values of the spectral parameter, namely $\lambda=\pm\omega/2$. To be explicit, we have
\begin{eqnarray}\label{eq:taupm}
    &&\tau\left(\frac{\omega}{2}\Big|\omega\right)={\cal U}\prod_{j=1}^N\left(\frac{\mathbb{1}+(-1)^{\delta_{j,N}}\tanh(\omega)\Gamma_{2j}\Gamma_{2j+1}}{1+{\rm i}\tanh(\omega)}\right),\nonumber\\
    &&\tau\left(-\frac{\omega}{2}\Big|\omega\right)={\cal U}\prod_{j=1}^N\left(\frac{\mathbb{1}-\tanh(\omega)\Gamma_{2j-1}\Gamma_{2j}}{1-{\rm i}\tanh(\omega)}\right),
\end{eqnarray}
where ${\cal U}$ is the twisted translation operator \eqref{eq:TwistTrans}. It is now straightforward to obtain
\begin{eqnarray}\label{eq:twisttrans2}
    {\cal U}^2=\tau\left(\frac{\omega}{2}\Big|\omega\right)\tau\left(-\frac{\omega}{2}\Big|\omega\right)=\tau\left(-\frac{\omega}{2}\Big|\omega\right)\tau\left(\frac{\omega}{2}\Big|\omega\right),
\end{eqnarray}
We note that, although we cannot get ${\cal U}$ from the inhomogeneous transfer matrix $\tau(\lambda|\omega),~\omega\neq 0$, we find ${\cal U}^2$ from the algebra generated by the inhomogeneous transfer matrices. Furthermore, since ${\cal U}^{2N}\propto{\cal P}$, we also have $\left[\tau\left({\omega}/{2}|\omega\right)\tau\left(-{\omega}/{2}|\omega\right)\right]^N\propto{\cal P}$. Therefore, the fermionic parity ${\cal P}$ remains a conserved quantity even for $\omega\neq0$. In particular, it commutes with all the other charges that can be derived from the transfer matrix. More importantly, since ${\cal U}^{4N}\propto\mathbb{1}$, this also establishes the existence of the operator $\left({\cal U}^2\right)^{-1}$ in the tower of the conserved charges.

Now one can immediately verify the following relations from \eqref{eq:TrottCirc} and \eqref{eq:taupm}:
\begin{eqnarray}\label{eq:transspacetime}
    \tau\left(\frac{\omega}{2}\Big|\omega\right)^2={\cal U}^2{\cal V}(\Omega),\quad \tau\left(-\frac{\omega}{2}\Big|\omega\right)^2={\cal U}^2{\cal V}(\Omega)^{\dagger},
\end{eqnarray}
with the identification $\Omega=\tanh(\omega)$. Combining the results from \eqref{eq:twisttrans2} and \eqref{eq:transspacetime}, the quantum circuit \eqref{eq:TrottCirc} finally can be obtained as
\begin{eqnarray}\label{eq:intcirc}
{\cal V}(\Omega) = \tau\left(-\frac{\omega}{2}\Big|\omega\right)^{-1}\tau\left(\frac{\omega}{2}\Big|\omega\right),\quad \Omega=\tanh(\omega).
\end{eqnarray}
This essentially proves the integrability of the trotterized circuit ${\cal V}(\Omega)$. Importantly, the time step $\Omega$ in the trotterized circuit is determined by the inhomogeneity parameter $\omega$. The first few local conserved quantities for this circuit are derived in Appendix \ref{app:LocCon}. Evidently, the conserved charges, $Q(\Omega)$'s, become functions of the time-step $\Omega$.

\subsection{From Majorana fermions to the critical TFIM}\label{sec:MajJWTFIM}
We now move to the spin-picture by representing the Majorana modes in terms of the spin-1/2 operators via the JW transformation 
\begin{eqnarray}
    \Gamma_{2j-1}=\left[\prod_{k=1}^{j-1}Z_k\right]X_j,~\Gamma_{2j}=\left[\prod_{k=1}^{j-1}Z_k\right]Y_j,
\end{eqnarray}
with $j=1,\cdots,N$. The integrable circuit from Equation \eqref{eq:TrottCirc} takes the form
\begin{eqnarray}
    &&{\cal V}(\Omega)=\prod\limits_{j=1}^{N}U^{\rm Z}_j(\Omega) \prod\limits_{j=1}^{N-1}U^{\rm XX}_j(\Omega) \left(\frac{{\mathbb 1}+{\rm i}\Omega {\sf P}X_NX_1}{1+{\rm i}\Omega}\right).
\end{eqnarray}
Here the unitaries $U^{\rm Z}_j(\Omega) ,U^{\rm XX}_j(\Omega) $ are as defined in Equation \eqref{eq:approxGates} and ${\sf P}=\prod_{j=1}^NZ_j=(-{\rm i})^N{\cal P}$ is the $\mathbb{Z}_2$ parity operator. However, although the above circuit is Yang-Baxter integrable, it also is highly non-local due to the presence of the operator ${\sf P}$, which has support over all the sites. Interestingly, we can isolate this non-local term in a clever way. We can rewrite the above circuit as
    \begin{eqnarray}
    &&{\cal V}(\Omega)={\sf V}(\Omega)-\frac{{\rm i}\Omega}{1+{\rm i}\Omega}\left[\prod_{j=1}^{N}U^{\rm Z}_j(\Omega) \prod_{j=1}^{N-1}U^{\rm XX}_j(\Omega)\right]\times\nonumber\\
    &&\hspace{5cm}({\mathbb 1}-{\sf P}){X_NX_1},
\end{eqnarray}
where 
\begin{eqnarray}\label{eq:FinalCircuit}
    {\sf V}(\Omega)&=&V_A(\Omega) V_B(\Omega)\nonumber\\
    &=&\prod_{j=1}^N\left(\frac{{\mathbb 1}+{\rm i}\Omega Z_j}{1+{\rm i}\Omega}\right)\prod_{j=1}^N\left(\frac{{\mathbb 1}+{\rm i}\Omega X_jX_{j+1}}{1+{\rm i}\Omega}\right),
\end{eqnarray}
is the required quantum circuit \eqref{eq:Gates} for the critical TFIM with the local quantum gates \eqref{eq:approxGates}. 
Now for every conserved charge $Q(\Omega) $, we consider another charge 
\begin{eqnarray}
  {\sf Q}(\Omega) :=\frac{1}{2}({\mathbb 1}+{\sf P})Q(\Omega)=\frac{1}{2}Q(\Omega)({\mathbb 1}+{\sf P}).  
\end{eqnarray}
This ${\sf Q}(\Omega) $ is a conserved quantity for the local, trotterized circuit ${\sf V}(\Omega)$ 
\begin{eqnarray}\label{eq:spinintcirc}
    \left[{\sf V}(\Omega) ,{\sf Q}(\Omega) \right]=0,\quad {\cal V}(\Omega)({\mathbb 1}+{\sf P})={\sf V}(\Omega)({\mathbb 1}+{\sf P}).
\end{eqnarray}
Here we used the orthogonality property of the projectors as $({\mathbb 1}+{\sf P})({\mathbb 1}-{\sf P})=0=({\mathbb 1}-{\sf P})({\mathbb 1}+{\sf P})$ and the fact that ${\sf P}$ commutes with both $X_NX_1$ and $\prod_{j=1}^{N}U^{\rm Z}_j(\Omega) \prod_{j=1}^{N-1}U^{\rm XX}_j(\Omega)$. We therefore have the set $\{{\sf Q}(\Omega) \}$, as the algebra of commuting observables, corresponding to the local quantum circuit ${\sf V}(\Omega)$. This completes the proof of the above quantum circuit being integrable. Notably, the conserved quantities we get are non-invertible by construction.

\section{Kramers-Wannier duality-symmetries of the trotterized critical Ising chain}\label{sec:DisKW}
In \cite{Sinha:2025wqf}, we showed that the non-invertible KW duality-symmetry operator for the critical Ising chain Hamiltonian can be obtained by the QISM formalism. The operator ${\sf D}$, which exchanges $H_A$ and $H_B$ in \eqref{eq:TFIM ham}, is a part of the abelian algebra generated by the commuting transfer matrices. In particular, it is given by \cite{Seiberg2024_NonInvertibleLSM,shao2023s}
    \begin{eqnarray}\label{eq:KWcont}
    {\sf D}&=&\frac{1}{2}{\cal U}({\mathbb 1}+{\sf P})\nonumber\\
    &=&{e^{-\frac{2\pi{\rm i}N}{8}}}\left(\prod_{j=1}^{N-1}\frac{{\mathbb 1}+{\rm i}Z_j}{\sqrt{2}}\frac{{\mathbb 1}+{\rm i}X_jX_{j+1}}{\sqrt{2}}\right)\frac{{\mathbb 1}+{\rm i}Z_N}{\sqrt{2}}\times\nonumber\\
    &&\hspace{5.5cm}\frac{({\mathbb 1}+{\sf P})}{2}.
\end{eqnarray}
where ${\cal U}$ is the twisted translation operator \eqref{eq:TwistTrans}. It is non-invertible by construction and thus cannot act by conjugation. Rather it acts as
\begin{eqnarray}\label{eq:KW}
    {\sf D}H_{A(B)}=H_{B(A)}{\sf D}.
\end{eqnarray} 
From here it is straightforward to see that it commutes with the critical TFIM Hamiltonian $[{\sf D},H_A+H_B]=0$, hence representing a non-invertible symmetry of the TFIM at the criticality.

In the trotterized case, however, the concept of the Hamiltonian generating an infinitesimal time evolution is lost. Rather we have the two unitaries $V_{A(B)}(\Omega)=\exp~(-{\rm i}H_{A(B)}\Omega)$, stroboscopically generating the time evolution. Therefore, the natural generalization of the usual KW duality is expected to map $V_{A(B)}(\Omega)\to V_{B(A)}(\Omega)$. It is easy to see that ${\sf D}$ itself does this job ${\sf D}V_{A(B)}(\Omega)=V_{B(A)}(\Omega){\sf D}$. Unfortunately, in the present scenario, ${\cal U}$ no longer can be obtained from the inhomogeneous transfer matrix $\tau(\lambda|\omega)$ with $\omega\neq 0$. Therefore, ${\sf D}$ no longer remains a symmetry of the trotterization, as can be seen from ${\sf D}{\sf V}(\Omega)=V_B(\Omega)V_A(\Omega){\sf D}$. We note that, although $V_B(\Omega)V_A(\Omega)$ approximates the continuous time evolution equally well, it is not the same as ${\sf V}(\Omega)$. This essentially stems from the fact that $H_A$ and $H_B$ do not commute with each other. Therefore the question arises: Are there conserved charges which possibly can implement the maps $V_{A(B)}(\Omega)\to V_{B(A)}(\Omega)$ and also commute with the unitary ${\sf V}(\Omega)$? {We now show that indeed there are such conserved charges, which can be constructed systematically from the transfer matrix. }

\subsection{KW duality and the non-invertible symmetries}\label{sec:DisKWDuSy}
Let us consider  the operators $\tau\left(\pm\frac{\omega}{2}\Big|\omega\right)$. The corresponding conserved charges in the spin-1/2 picture are obtained by multiplying the factor ${({\mathbb 1}+{\sf P})}/{2}$ as
\begin{eqnarray}
    {\frak D}_\pm(\Omega) :=\frac{1}{2}\tau\left(\pm\frac{\omega}{2}\Big|\omega\right)({\mathbb 1}+{\sf P}),\quad \Omega=\tanh(\omega).
\end{eqnarray}
By the very construction, they commute with the time evolution $[{\frak D}_{\pm}(\Omega),{\sf V}(\Omega)]=0$. To be explicit, they have the following expressions in the spin language:
\begin{eqnarray}\label{eq:DiscreteKWexp}
    &&{\frak D}_-(\Omega)={\sf D}\prod_{j=1}^N\left(\frac{{\mathbb 1}-{\rm i}\Omega Z_j}{1-{\rm i}\Omega}\right),\nonumber\\
    &&{\frak D}_+(\Omega)={\sf D}\prod_{j=1}^N\left(\frac{{\mathbb 1}+{\rm i}\Omega X_jX_{j+1}}{1+{\rm i}\Omega}\right),
\end{eqnarray}
with ${\sf D}$ given in \eqref{eq:KWcont}. Both ${\frak D}_{\pm}(\Omega)$ go to the well-known non-invertible KW duality-symmetry operator ${\sf D}$ if we take the homogeneous limit $\lim_{\Omega\to0}{\frak D}_\pm(\Omega)={\sf D}$. As before, these charges are non-invertible. We define their action as 
\begin{eqnarray}
    {\frak D}_{\pm}(\Omega):{\cal O}\to {\cal O}_\pm,~~\text{with}~~ {\frak D}_\pm(\Omega){\cal O}={\cal O}_\pm{\frak D}_\pm(\Omega).
\end{eqnarray}
In Table \ref{tab:actionDpm}, we summarize how they act on some relevant unitaries.
\begin{table}[h!]
    \centering
    \begin{tabular}{c c c}
    \hline
    \hline
        \boldsymbol{${\cal O}$} & \boldsymbol{${\cal O}_-$} & \boldsymbol{${\cal O}_+$} \\
        \hline
        $V_A(\Omega)$ & $V_B(\Omega)$ & $V_A(\Omega)V_B(\Omega)V_A(\Omega)^{\dagger}$ \\
        $V_B(\Omega)$ & $V_B(\Omega)^{\dagger}V_A(\Omega)V_B(\Omega)$ & $V_A(\Omega)$\\
        $V_A(\Omega)V_B(\Omega)$ & $V_A(\Omega)V_B(\Omega)$ & $V_A(\Omega)V_B(\Omega)$\\
        \hline
        \hline
    \end{tabular}
    \caption{The action of the non-invertible operators ${\frak D}_\pm$ on the unitaries $V_A,V_B,V_AV_B$.}
    \label{tab:actionDpm}
\end{table}\\
Therefore, we found the operators that implement the transformation  $V_{A(B)}\to V_{B(A)}$ and also commute with the time evolution 
\begin{eqnarray}
    &&{\frak D}_{-(+)}(\Omega)V_{A(B)}(\Omega)=V_{B(A)}(\Omega){\frak D}_{-(+)}(\Omega),\nonumber\\
    &&\left[{\frak D}_\pm(\Omega),{\sf V}(\Omega)\right]=\left[{\frak D}_\pm(\Omega),V_A(\Omega)V_B(\Omega)\right]=0.
\end{eqnarray}
Note that, the non-invertible operators act quite differently on $V_A$ and $V_B$. This is crucial to ensure the commutativity between the symmetry operators and the trotterized time evolution.

Let us now investigate how ${\frak D}_\pm(\Omega)$ implement the duality transformation for the trotterized dynamics of the generic TFIM. Consider the Hamiltonian
\begin{eqnarray}\label{eq:GenTFIMHam}
    {\sf H}_{\rm TFIM}(h,J)
    &=&-h\sum_{j=1}^NZ_j-J\sum_{j=1}^NX_jX_{j+1}\nonumber\\
    &=&hH_A+JH_B
\end{eqnarray}
which becomes the critical TFIM \eqref{eq:TFIM ham} at $h=1=J$. The continuous-time KW duality-symmetry operator ${\sf D}$, as defined in \eqref{eq:KWcont}, interchanges the above Hamiltonian and its dual one as ${\sf D}{\sf H}_{\rm TFIM}(h,J)={\sf H}_{\rm TFIM}(J,h){\sf D}$. We take the corresponding quantum circuit as 
\begin{eqnarray}\label{eq:GenTFIMDisTime}
    {\sf V}(\Omega;h,J)&=&\prod_{j=1}^N\left(\frac{{\mathbb 1}+{\rm i}h\Omega Z_j}{1+{\rm i}h\Omega}\right)\prod_{j=1}^N\left(\frac{{\mathbb 1}+{\rm i}J\Omega X_jX_{j+1}}{1+{\rm i}J\Omega}\right)\nonumber\\
    &=&V_A(h\Omega)V_B(J\Omega).
\end{eqnarray}
which approximates the time evolution $\exp[-{\rm i}t{\sf H}_{\rm TFIM}(h,J)]\simeq{\sf V}(\Omega;h,J)^n+{\cal O}(t^2/n)$, with $\Omega=t/n$. We expect the operators ${\frak D}_\pm(\Omega)$ to implement some duality transformation on the above circuit. It turns out that, their actions differ depending on how we describe the circuit itself. To be precise, we have:
\begin{eqnarray}\label{eq:DualDisTFIM}
    &&{\frak D}_-(\Omega){\sf V}(\Omega;1,J)={\sf V}(\Omega;J,1){\frak D}_-(\Omega),\label{eq:D_}\\
    &&{\frak D}_+(\Omega){\sf V}(\Omega;h,1)={\sf V}(\Omega;1,h){\frak D}_+(\Omega).\label{eq:D+}
\end{eqnarray}
We set $h=1$ and $J=1$ in Equations \eqref{eq:D_} and \eqref{eq:D+}, respectively. This essentially ensures that ${\frak D}_\pm(\Omega)$, acting on ${\sf V}(\Omega;h,J)$, do not change the order of the unitaries $V_A$ and $V_B$. The above choice of the quantum circuit guarantees that at the criticality $J=1=h$, the unitary ${\sf V}(\Omega;1,1)={\sf V}(\Omega)$ commutes with the non-invertible duality operators ${\frak D}_\pm(\Omega)$ as $[{\sf V}(\Omega;1,1),{\frak D}_\pm(\Omega)]=0$. Therefore, we found the desired discrete-time KW duality operators ${\frak D}_\pm(\Omega)$, which implement the necessary duality transformations, as well as become a non-invertible symmetry of the trotterized critical TFIM.

\subsection{Algebra of the non-invertible ${\frak D}_\pm(\Omega)$}\label{sec:AlgDisKW}

\begin{figure}[h!]
        \centering
        \includegraphics[width=1\linewidth]{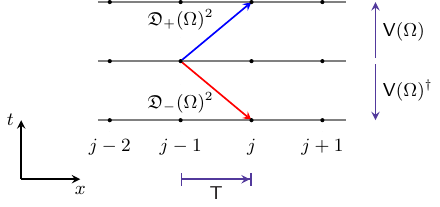}
        \caption{We can think of ${\frak D}_\pm(\Omega)^2$ as translations along the light-cone coordinates $x\pm t$. We restricted ourselves in the even parity sector ${\sf P}=1$.}
        \label{fig:halftranslation}
\end{figure}

We now determine the algebra generated by the non-invertible operators ${\frak D}_\pm(\Omega)$. We begin by recalling the well-known algebra satisfied by the continuous-time KW duality-symmetry operator \cite{Seiberg2024_NonInvertibleLSM,shao2023s}
\begin{eqnarray}\label{eq:KWalg}
    {\sf D}^2=\frac{1}{2}({\mathbb 1}+{\sf P}){\sf T},\qquad{\sf D}^\dagger{\sf D}=\frac{1}{2}({\mathbb 1}+{\sf P}),
\end{eqnarray}
where ${\sf T}=P_{1,2}\cdots P_{N-1,N}$ is the translation operator that shifts the spin indices by one, with $P_{j,k}=(\mathbb{1}+X_jX_k+Y_jY_k+Z_jZ_k)/2$ being the permutation operator which exchanges the indices $j$ and $k$. To see this, we can consider the parity even subalgebra generated by $\{Z_j\}$ and $\{X_jX_{j+1}\}$. It can be checked that $(1+{\sf P}){\cal U}^2$ and $(1+{\sf P}){\sf T}$ has the same action on the above subalgebra. Often the operator ${\sf D}$ is therefore regarded as the ``half spatial translation". Now let us consider trotterized KW duality operators ${\frak D}_\pm(\Omega)$. From \eqref{eq:transspacetime} and \eqref{eq:spinintcirc}, it is straightforward to establish
\begin{eqnarray}\label{eq:halfspacetime}
    &&{\frak D}_+(\Omega)^2=\frac{1}{2}({\mathbb 1}+{\sf P}){\sf T}{\sf V}(\Omega),\nonumber\\
    &&{\frak D}_-(\Omega)^2=\frac{1}{2}({\mathbb 1}+{\sf P}){\sf T}{\sf V}(\Omega)^{\dagger}.
\end{eqnarray}
Therefore, the algebra generated by the Kramers-Wannier duality operators in the trotterized case involves spatial translation as well as time evolution operations. In particular, ${\frak D}_\pm(\Omega)$ now can be regarded as ``half spatio-temporal translation" operators (Fig. \ref{fig:halftranslation}). Alternatively, ${\frak D}_\pm(\Omega)$ can also be interpreted as ``half-translation" along the light-cone coordinates $x\pm t$ \cite{destri1987light,faddeev1996algebraic}. 
This essentially stems from the fact that in contrast to the continuous time case, where the space and time are not quite on equal footing, when we trotterize the circuit, both the space and time are discretized and hence are on equal footing.
They further satisfy
\begin{eqnarray}\label{eq:GenKWAlg}
    &&{\frak D}_+^\dagger(\Omega){\frak D}_+(\Omega)=\frac{1}{2}({\mathbb 1}+{\sf P})={\frak D}_-^\dagger(\Omega){\frak D}_-(\Omega),\nonumber\\
    &&{\frak D}_-^\dagger(\Omega){\frak D}_+(\Omega)=\frac{1}{2}({\mathbb 1}+{\sf P}){\sf V}(\Omega),\nonumber\\
    &&{\frak D}_\pm(\Omega){\frak D}_\mp(\Omega)=\frac{1}{2}({\mathbb 1}+{\sf P}){\sf T}.
\end{eqnarray}
As one can see, the algebra includes both spatial and temporal translation operators. At this stage, it is tempting to consider the $\Omega\to 0$ limit. Indeed for infinitesimal $\Omega$, we have
\begin{eqnarray}
    &&{\frak D}_\pm(\Omega)\simeq {\frak D}(0)+\Omega\frac{\partial{\frak D}_\pm(\Omega)}{\partial{\Omega}}\Big|_{\Omega=0},\nonumber\\
    &&{V}_{A(B)}(\Omega)\simeq {V}_{A(B)}(0)+\Omega\frac{\partial V_{A(B)}}{\partial{\Omega}}\Big|_{\Omega=0}.
\end{eqnarray}
With ${\frak D}(0)={\sf D},{V}_{A(B)}(0)=\mathbb{1},\partial{V}_{A(B)}(\Omega)/\partial{\Omega}|_{\Omega=0}=-{\rm i}H_{A(B)}$, we recover \eqref{eq:KWalg} from \eqref{eq:halfspacetime},\eqref{eq:GenKWAlg} and \eqref{eq:KW} from Table \eqref{tab:actionDpm}.

\section{Kramers-Wannier duality-symmetry for a Floquet TFIM}\label{sec:KWFloquet}
So far we talked about the integrable trotterization of the critical TFIM, where the trotterized quantum circuit was given by \eqref{eq:Gates}. To lower the Trotter error, we keep the time-step sufficiently small, enabling the approximation \eqref{eq:approxGates}. Among the other conserved quantities, we found the non-invertible operators ${\frak D}_\pm$ which can be regarded as implementing the well-known KW duality in the trotterized TFIM. We now present a broader perspective of the discrete-time dynamics in which the time-step is not required to be small. Evidently a significant Trotter error can occur for a sufficiently big time-step, making it difficult to provide an accurate approximation for the continuous-time evolution. Nonetheless, such an evolution can still be conceptualized as a two-step Floquet evolution with a finite time period.
\subsection{Integrable Floquet TFIM}
To begin with, we briefly discuss some aspects of a two-step Floquet evolution. This essentially is a periodic driving protocol, where the time evolution over a complete time period $2t$ is composed of distinct, sequential Hamiltonian evolutions, each with a time-period of duration $t$. To be precise, we shall work with the Floquet operator \cite{gritsev2017integrable}
\begin{eqnarray}
    {\sf V}^{\cal F}(t;h,J)=e^{-{\rm i}ht H_A}e^{-{\rm i}Jt H_B},
\end{eqnarray}
where $t$ is not necessarily small and the Hamiltonians $H_A,H_B$ are defined in \eqref{eq:TFIM ham}. The effective Floquet Hamiltonian $H^{\cal F}$ is obtained by taking the logarithmic of the evolution operator as $-2{\rm i}tH^{\cal F}=\log\left[{\sf V}^{\cal F}(t;h,J)\right]$. 
The above circuit comes with an extensive set of conserved quantities, which can be obtained by exploiting the well-known \textit{Onsager algebra} \footnote{Interestingly, our construction using the inhomogeneous transfer matrix can generate the entire Onsager algebra as well. We provide the relevant details in Appendix \ref{app:Onsager}.}, as shown in \cite{prosen1998new,Prosen2000ExactTF,gritsev2017integrable}. 
Interestingly, as long as $|t|\leq\pi/4$ holds, we can obtain the circuit ${\sf V}^{\cal F}(t;1,1)$ from the transfer matrix \eqref{eq:inhomMonoIsing}, thus making it Yang-Baxter integrable as well. To see this, we expand the above circuit as
\begin{eqnarray}
    &&{\sf V}^{\cal F}(t;1,1)=\prod_{j=1}^N\left(\cos(t){\mathbb 1}+{\rm i}\sin(t)Z_j\right)\times\nonumber\\
    &&\hspace{2cm}\prod_{j=1}^N\left(\cos(t){\mathbb 1}+{\rm i}\sin(t)X_jX_{j+1}\right).
\end{eqnarray}
with the periodic boundary condition $N+1\equiv 1$. Here we made use of the fact that $(Z_j)^2={\mathbb 1}=(X_jX_{j+1})^2$. Now if we consider the integrable circuit \eqref{eq:FinalCircuit} and make the identification $\tan(t)=\Omega$, we essentially have
\begin{eqnarray}
    {\sf V}^{\cal F}(t;1,1)=e^{2{\rm i}Nt}{\sf V}(\Omega).
\end{eqnarray}
Since we already established the integrability of the circuit \eqref{eq:FinalCircuit}, the above identification renders the unitary ${\sf V}^{\cal F}(t;1,1)$ also to be integrable. Furthermore, the time-step $\Omega$ is related to the inhomogeneity parameter $\omega$ through the relation $\Omega=\tanh(\omega)$. This restricts the range of the values that $\Omega$ can assume as $-1\leq\Omega\leq1$. Combining the above results, we finally obtain the required relation for the integrability as
\begin{eqnarray}
    -\frac{\pi}{4}\leq t\leq \frac{\pi}{4}.
\end{eqnarray}
This completes our required proof.

\subsection{KW duality-symmetry of the Floquet evolution}
We will now identify the KW duality for the above Floquet evolution. Let us define the non-invertible operators ${\frak D}_\pm^{\cal F}(t)$ as
\begin{eqnarray}\label{eq:QCGEN}
    {\frak D}_-^{\cal F}(t)={\sf D}e^{{\rm i}tH_A},\quad {\frak D}_+^{\cal F}(t)={\sf D}e^{-{\rm i}tH_B},
\end{eqnarray}
with ${\sf D}$ being the non-invertible, continuous-time KW duality-symmetry operator \eqref{eq:KWcont}. Again, as long as $\tan(t)=\Omega$, we have
\begin{eqnarray}
    {\frak D}_-^{\cal F}(t)=e^{-{\rm i}Nt}{\frak D}_-(\Omega),\quad {\frak D}_+^{\cal F}(t)=e^{{\rm i}Nt}{\frak D}_+(\Omega).
\end{eqnarray}
These non-invertible operators act on the circuit ${\sf V}^{\cal F}(t;h,J)$ in the following ways:
\begin{eqnarray}\label{eq:GeneralDuality}
    &&{\frak D}_-^{\cal F}(t){\sf V}^{\cal F}(t;h,J)=e^{-{\rm i}(h-1)tH_B}{\sf V}^{\cal F}(t;J,1){\frak D}_-^{\cal F}(t),\\
    &&{\frak D}_+^{\cal F}(t){\sf V}^{\cal F}(t;h,J)={\sf V}^{\cal F}(t;1,h)e^{-{\rm i}(J-1)tH_A}{\frak D}_+^{\cal F}(t).
\end{eqnarray}
Clearly, setting $h=1$ and $J=1$ in the respective equations, we obtain
\begin{eqnarray}
    &&{\frak D}_-^{\cal F}(t){\sf V}^{\cal F}(t;1,J)={\sf V}^{\cal F}(t;J,1){\frak D}_-^{\cal F}(t),\nonumber\\
    &&{\frak D}_+^{\cal F}(t){\sf V}^{\cal F}(t;h,1)={\sf V}^{\cal F}(t;1,h){\frak D}_+^{\cal F}(t).
\end{eqnarray}
Now it is straightforward to check that if $t$ is sufficiently small such that $\Omega=\tan(t)\simeq t$, we arrive at \eqref{eq:D_} and \eqref{eq:D+}. Another interesting situation arises if we set $h=2$ and $J=2$ in the respective equations. In this scenario, we have the relations
\begin{eqnarray}\label{eq:FloquetDW2}
    &&{\frak D}_-^{\cal F}(t){\sf V}^{\cal F}(t;2,J)=\left[e^{-{\rm i}tH_B}e^{-{\rm i}JtH_A}e^{-{\rm i}tH_B}\right]{\frak D}_-^{\cal F}(t),\nonumber\\
    &&{\frak D}_+^{\cal F}(t){\sf V}^{\cal F}(t;h,2)=\left[e^{-{\rm i}tH_A}e^{-{\rm i}htH_B}e^{-{\rm i}tH_A}\right]{\frak D}_+^{\cal F}(t).
\end{eqnarray}
The expressions inside the braces on the right hand side can be regarded as some three-step Floquet evolutions. However, it is the small $t$ limit which is much more interesting. Let us define 
\begin{eqnarray}
    &&{\sf V}^{\cal F}_-(t;h,J)=e^{-{{\rm i}Jt  H_B}/{2}}e^{-{\rm i}ht H_A}e^{-{{\rm i}Jt H_B}/{2}},\nonumber\\
    &&{\sf V}^{\cal F}_+(t;h,J)=e^{-{{\rm i}ht  H_A}/{2}}e^{-{\rm i}Jt H_B}e^{-{{\rm i}ht H_A}/{2}}.
\end{eqnarray}
By consecutively applying the above operators, we can once again approximate the continuous-time evolution $e^{-{\rm i}t{\sf H}_{\rm TFIM}(h,J)}$ as
\begin{eqnarray}
    \left[{\sf V}^{\cal F}_-(\Omega;h,J)\right]^n\simeq e^{-{\rm i}t{\sf H}_{\rm TFIM}(h,J)}\simeq \left[{\sf V}^{\cal F}_+(\Omega;h,J)\right]^n,
\end{eqnarray}
with $\Omega=t/n$. Often, this particular approximation is dubbed as the \textit{second-order trotterization}. It can be regarded as an improvement over the usual trotterization \eqref{eq:1stTrot}, in the sense that the Trotter error now becomes ${\cal O}(t^3/n^2)$ (e.g. see Chapter $4$.\textit{Quantum Circuits} from \cite{nielsen2010quantum}). Crucially, all the circuits, ${\sf V}^{\cal F},{\sf V}^{\cal F}_-,{\sf V}^{\cal F}_+$, approximate the same continuous-time evolution. We now can show that, the operators ${\frak D}_\pm^{\cal F}(\Omega)$ also map the usual trotterized circuit ${\sf V}^{\cal F}$ to the dual, second-order trotterized circuits ${\sf V}_{\pm}^{\cal F}$ as
\begin{eqnarray}\label{eq:1stTrott2ndTrott}
    &&{\frak D}_-^{\cal F}(\Omega){\sf V}^{\cal F}(\Omega;2,J)={\sf V}^{\cal F}_-(\Omega;J,2){\frak D}_-^{\cal F}(\Omega),\nonumber\\
    &&{\frak D}_+^{\cal F}(\Omega){\sf V}^{\cal F}(\Omega;h,2)={\sf V}^{\cal F}_+(\Omega;2,h){\frak D}_+^{\cal F}(\Omega).
\end{eqnarray}
It, however, should be emphasized that we did not derive the circuits ${\sf V}_{\pm}^{\cal F}$ by our transfer matrix formalism. Therefore, they cannot be claimed as Yang-Baxter integrable. Nonetheless, they appear as the duality transformed circuits under the action of the relevant KW duality-symmetry operators.

{
\section{Phase structure}\label{sec:phase-structure}
As an immediate application of our construction, we now examine some consequences of the non-invertible duality operators ${\frak D}_\pm(\Omega)$ for the phase structure of the trotterized quantum circuit. Later we shall extend the discussion to include the Floquet phases as well.
\subsection{Trotterized case}
The circuit ${\sf V}(\Omega;h,J)$ being unitary, comes with the spectrum ${\sf V}(\Omega;h,J)|E_{(h,J)}(\Omega)\rangle=e^{{\rm i}E_{(h,J)}(\Omega)}|E_{(h,J)}(\Omega)\rangle$, where $\{E_{(h,J)}(\Omega)\}$ are the \textit{quasienergies}, defined modulo $2\pi$. From \eqref{eq:DualDisTFIM}, it is now easy to check that, the non-invertible operator ${\frak D}_+(\Omega)$ maps a parity-even eigenstate ${\sf P}|E_{(h,1)}(\Omega)\rangle=|E_{(h,1)}(\Omega)\rangle$ to its dual as 
\begin{eqnarray}\label{eq:noninvacteigen}
    {\frak D}_+(\Omega)|E_{(h,1)}(\Omega)\rangle=|E_{(1,h)}(\Omega)\rangle,
\end{eqnarray}
with the same quasienergy $E_{(1,h)}(\Omega)=E_{(h,1)}(\Omega)$. As one can verify, the resulting state $|E_{(1,h)}(\Omega)\rangle$ also has even parity. It remains to say that one can use ${\frak D}_-(\Omega)$ equally well. Therefore, the non-invertible duality operators essentially fix the spectrum of the dual quantum circuit in the even parity sector. However, since ${\frak D}_\pm(\Omega)$ annihilate all the parity-odd states, such a statement cannot be made for the odd parity subspace. Indeed, Fig.~\ref{fig:evenoddspectra} shows that the spectra in this sector differ.
\begin{figure}[h!]
    \centering
    \includegraphics[width=1\linewidth]{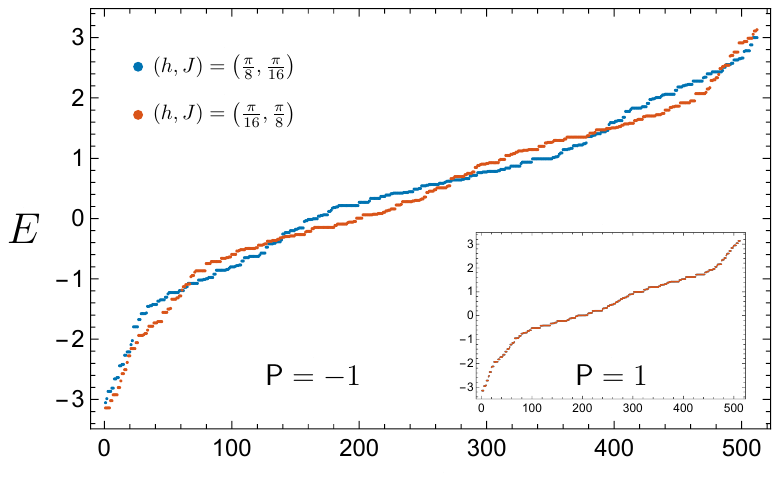}
    \caption{The quasi energies $-\pi\leq E\leq\pi$ of the Floquet evolution ${\sf V}(1;h,J)$ in \eqref{eq:GenTFIMDisTime} are plotted for $N=10$ spins. We take the parameter values $(h,J)\in(\pi/8,\pi/16)$. Since the dual circuits ${\sf V}(1;h,J)$ and ${\sf V}(1;J,h)$ are intertwined by the non-invertible KW duality-symmetries ${\frak D}_-(h)$ and ${\frak D}_+(J)$ in \eqref{eq:DiscreteKWexp}, they share a common spectrum in the even-parity $({\sf P}=1)$ sector. However, the spectra are clearly different in the parity-odd $({\sf P}=-1)$ subspace.}
    \label{fig:evenoddspectra}
\end{figure}

\subsubsection{Non-existence of invertible extension}
We now show that, the non-invertible dualites ${\frak D}_\pm(\Omega)$ cannot be extended to the odd parity subspace. As a result, the spectra of the dual Hamiltonians in the odd parity subspace do not have a simple relation. To begin with, note that the local quantum circuit ${\sf V}(\Omega)$ in \eqref{eq:Gates} can also be written as 
\begin{eqnarray}
    {\sf V}(\Omega)={\cal V}(\Omega){\sf P}_++\tilde{\cal V}(\Omega){\sf P}_-,
\end{eqnarray}
where $\tilde{\cal V}(\Omega)=\prod_j\tilde{U}_{2j-1,2j}(\Omega)\prod_j\tilde{U}_{2j,2j+1}(\Omega)$, is the quantum circuit for the periodic Majorana chain $\tilde{H}={\rm i}\sum_j\Gamma_j\Gamma_{j+1}$, with the local quantum gate $\tilde{U}_{j,j+1}(\Omega)\simeq(\mathbb{1}+\Omega\Gamma_j\Gamma_{j+1})/(1+{\rm i}\Omega)$. We further defined ${\sf P}_\pm=({\mathbb 1}\pm{\sf P})/2$ as the projectors to the even and odd parity sectors, respectively. This reproduces the familiar result that periodic (antiperiodic) boundary condition in the spin-1/2 chain translates into antiperiodic (periodic) boundary condition in the corresponding fermionic description.
Recall that ${\frak D}_+(\Omega)$ is the projection of the transfer matrix $\tau(\omega/2|\omega)$ onto the even parity subspace, where $\tau(\omega/2|\omega)={\cal U}\left(\prod_{j=1}^NU_{2j,2j+1}(\Omega)\right)$ generates the half-translation along the light-cone coordinate $x+t$ \eqref{eq:transspacetime}. Therefore, a natural extension of the non-invertible duality can be obtained as ${\frak D}_+(\Omega)+\tilde{\frak D}_+(\Omega)$, where $\tilde{\frak D}_+(\Omega)$ is the projection of the light-cone half-translation operator of the periodic chain onto the odd-parity sector. Considering $\tilde{\cal U}:\Gamma_{j}\to\Gamma_{j+1}$ as the translation symmetry of the periodic Majorana chain, we can now show that the operator $\tilde{\cal U}\left(\prod_{j=1}^N\tilde{U}_{2j,2j+1}(\Omega)\right)$ squares to $\tilde{\cal U}^2 \tilde{\cal V}(\Omega)$ and thus serves as the appropriate half-translation operator for the time-evolution $\tilde{\cal V}(\Omega)$. Hence, we essentially have
\begin{eqnarray}
    \tilde{\frak D}_+(\Omega)={\sf P}_-\tilde{\cal U}\left(\prod_{j=1}^N\tilde{U}_{2j,2j+1}(\Omega)\right){\sf P}_-.
\end{eqnarray}
As can be checked further, the actions of ${\cal U}\left(\prod_{j=1}^NU_{2j,2j+1}(\Omega)\right)$ and $\tilde{\cal U}\left(\prod_{j=1}^N\tilde{U}_{2j,2j+1}(\Omega)\right)$ are the same in the bulk of the chain and differ only at the boundary. This reflects the fact that the discrete-time evolutions ${\cal V}(\Omega)$ and $\tilde{\cal V}(\Omega)$ also differ only through a boundary term. It is now trivial to verify that $\tilde{\cal U}\left(\prod_{j=1}^N\tilde{U}_{2j,2j+1}(\Omega)\right)$ commutes with the circuit $\tilde{\cal V}(\Omega)$. Therefore, ${\frak D}_+(\Omega)+\tilde{\frak D}_+(\Omega)$ indeed is a symmetry of the spin-1/2 evolution ${\sf V}(\Omega)$. However, the one-site translation symmetry of a periodic Majorana chain always anticommutes with the parity $\{\tilde{\cal U},{\sf P}\}=0$, as seen from \cite{hsieh2016all}
\begin{eqnarray}
    \tilde{\cal U}:\prod_{j=1}^{2N}\Gamma_j\to\left(\prod_{j=2}^{2N}\Gamma_j\right)\Gamma_1=-\prod_{j=1}^{2N}\Gamma_j.
\end{eqnarray}
This essentially leads to $\tilde{\frak D}_+(\Omega)=0$. In a completely analogous fashion, one can show that $\tilde{\frak D}_-(\Omega)$ also vanishes. Therefore, we fail to promote ${\frak D}_\pm(\Omega)$ to invertible operators. This explains the spectral mismatch in the odd-parity sector.

\subsubsection{Order-disorder duality}
We now demonstrate that ${\frak D}_\pm(\Omega)$ induce order–disorder duality within the parity-even eigenspace of the circuit ${\sf V}(\Omega;h,J)$. With a little effort, one can show that the non-invertible operator ${\frak D}_\pm(\Omega)$ has the following actions on the local terms:
\begin{eqnarray}\label{eq:KWactions}
    &&{\frak D}_-(\Omega)Z_j=X_jX_{j+1}{\frak D}_-(\Omega),\nonumber\\
    &&{\frak D}_+(\Omega)X_jX_{j+1}=Z_{j+1}{\frak D}_+(\Omega).
\end{eqnarray}
On the contrary, ${\frak D}_\pm(\Omega)$ do not have such simple actions on the local terms $Z_j$ and $X_jX_{j+1}$, respectively. Let us now consider the parity preserving order correlation $\langle X_jX_k\rangle$, with $j<k$, in the parity-even eigenstate $|E_{(h,1)}(\Omega)\rangle$ as $\langle X_jX_k\rangle_{h,1}=\langle E_{(h,1)}(\Omega)|X_jX_k|E_{(h,1)}(\Omega)\rangle$. Using the relations \eqref{eq:noninvacteigen},\eqref{eq:KWactions} and recalling that ${\frak D}_+(\Omega)^\dagger{\frak D}_+(\Omega)=(\mathbb{1}+{\sf P})/2$, one then arrives at 
\begin{eqnarray}\label{eq:order-disorder exp}
    &&\langle X_jX_k\rangle_{h,1}=\langle Z_{j+1}\cdots Z_{k} \rangle_{1,h},
\end{eqnarray}
with the disorder correlation $\langle Z_{j+1}\cdots Z_{k} \rangle_{1,h}=\langle E_{(1,h)}(\Omega)|Z_{j+1}\cdots Z_{k}|E_{(1,h)}(\Omega)\rangle$. Therefore, the non-invertible duality operator ${\frak D}_+(\Omega)$ ensures that the order correlations at $(h,1)$ equals to the disorder correlations at the dual couplings $(1,h)$. Subsequently the two sides of the critical line $h=1$ must describe two entirely different phases \cite{khemani2016phase}. Moreover, it confirms that at $h=1$, where the order and disorder correlations coincide, the system becomes self-dual and the non-invertible duality becomes a symmetry of the quantum circuit $[{\frak D}_+(\Omega),{\sf V}(\Omega;1,1)]=0$. A similar discussion involving the non-invertible duality ${\frak D}_-(\Omega)$ leads to the same conclusion. 

We emphasize that, although the continuous-time KW duality operator ${\sf D}$ relates the order and disorder operators, it does not map the quantum circuit ${\sf V}(\Omega;h,J)$ to its dual ${\sf V}(\Omega;J,h)$. Consequently, we cannot use ${\sf D}$ to arrive at the relation \eqref{eq:order-disorder exp}. Therefore, it is the non-invertible operators ${\frak D}_\pm(\Omega)$ that ensure the duality for the trotterized circuit ${\sf V}(\Omega;h,J)$.

\subsection{Floquet case}
We now concentrate on the Floquet evolution
\begin{eqnarray}\label{eq:Floquet2nd}
    {\sf V}^{\cal F}(h,J)=e^{-{\rm i}h H_A}e^{-{\rm i}J H_B},
\end{eqnarray}
where the time interval $t$ has been absorbed in the couplings $(h,J)$. As explained in \cite{khemani2016phase,PhysRevB.93.245145,von2016phase}, it suffices to restrict our analysis to the region $(h,J)\in[0,\pi/2]\times[0,\pi/2]$. The Floquet KW duality operators ${\frak D}^{\cal F}_\pm(x)$, defined in \eqref{eq:QCGEN}, act as
\begin{eqnarray}\label{eq:dual1}
    &&{\frak D}^{\cal F}_-(h){\sf V}^{\cal F}(h,J)={\sf V}^{\cal F}(J,h){\frak D}^{\cal F}_-(h),\nonumber\\
    &&{\frak D}^{\cal F}_+(J){\sf V}^{\cal F}(h,J)={\sf V}^{\cal F}(J,h){\frak D}^{\cal F}_+(J).
\end{eqnarray}
\begin{figure}[H]
    \centering
    \includegraphics[width=1\linewidth]{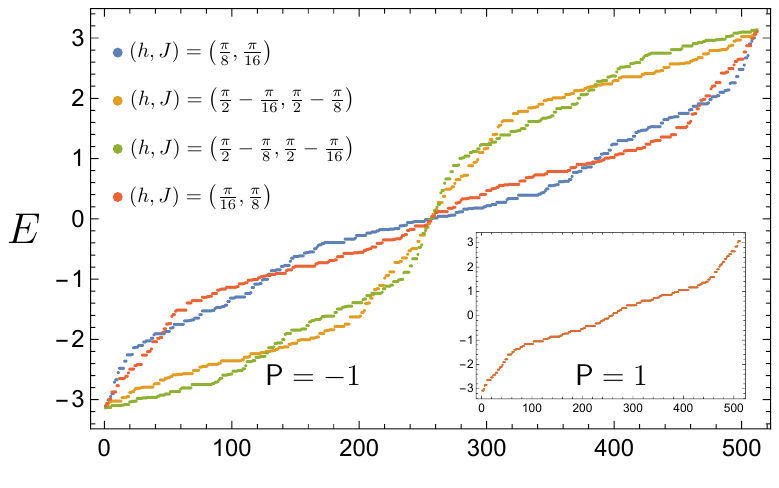}
    \caption{The quasi-energies for the Floquet evolution ${\sf V}^{\cal F}(h,J)$ are plotted for $N=10$. Specifically, we choose four different set of parameters $(h,J)_{\rm I}=(h_0,J_0)=(\pi/8,\pi/16),\,(h,J)_{\rm II}=(\pi/2-J_0,\pi/2-h_0),\,(h,J)_{\rm III}=(\pi/2-h_0,\pi/2-J_0)$ and $\,(h,J)_{\rm IV}=(J_0,h_0)$. Since the above four couplings are related by the non-invertible mappings ${\frak D^{\cal F}_\pm,\frak R^{\cal F}_\pm}$, their spectra coincide exactly in the even-parity sector. On the other hand, the spectra are distinct in the odd-parity subspace.}
    \label{fig:floquetspectra}
\end{figure}
As before, we can show that they cannot be extended to the parity-odd subspace. At $h=J$, the evolution commutes with ${\frak D}^{\cal F}_\pm$ and becomes self-dual under the above duality  transformations. Subsequently, the line $h=J$ represents a critical line separating the $h<J$ and the $h>J$ phases. This is the discrete version of the KW duality of the quantum Ising chain. However, the phase structure of the Floquet evolution ${\sf V}^{\cal F}(h,J)$ turns out to be much richer than its continuous-time counterpart \cite{khemani2016phase,PhysRevB.93.245145,von2016phase,berdanier2018floquet}. In particular, it possesses another critical line given by $h+J=\pi/2$. We now explicitly construct the duality operators, which map parameters across this critical line. To be precise, they map the couplings $(h,J)$ to $(\pi/2-J,\pi/2-h)$. 
The quasienergies of the Floquet evolution are plotted in Fig. \ref{fig:floquetspectra}.

\subsubsection{Additional non-invertible duality-symmetries}
To begin with, we introduce the ${\mathbb Z}_2$ operators ${\frak X}=\prod_{j=1}^NX_j$ and ${\frak Z}=\prod_{j=1}^{N/2}Z_{2j}$, with ${\frak X}^2={\mathbb 1}={\frak Z}^2$. Throughout, we assume $N$ to be even. It is then easy to check that both ${\frak X}$ and ${\frak Z}$ preserve parity $[{\frak X},{\sf P}]=0=[{\frak Z},{\sf P}]$. Furthermore, they satisfy $\{{\frak X},H_A\}=0=\{{\frak Z},H_B\}$ and $[{\frak X},H_B]=0=[{\frak Z},H_A]$. 
Let us now introduce two new non-invertible operators as
\begin{eqnarray}
    &&{\frak R}^{\cal F}_+(x)=e^{\frac{2\pi{\rm i} N}{8}}{\frak X}{\frak D}^{\cal F}_+(x){\frak X},\nonumber\\
    &&{\frak R}^{\cal F}_-(x)=e^{-\frac{2\pi{\rm i} N}{8}}{\frak Z}{\frak D}^{\cal F}_-(x){\frak Z}.
\end{eqnarray}
Crucially, they commute with the parity $[{\frak R}_\pm^{\cal F}(x),{\sf P}]=0$. With a little effort, one can now show that these operators act on the Floquet evolution as
\begin{eqnarray}\label{eq:dual2}
    &&{\frak R}^{\cal F}_+(J){\sf V}^{\cal F}(h,J)={\sf V}^{\cal F}\left(\frac{\pi}{2}-J,\frac{\pi}{2}-h\right){\frak R}^{\cal F}_+(J),\nonumber\\
    &&{\frak R}^{\cal F}_-(h){\sf V}^{\cal F}(h,J)={\sf V}^{\cal F}\left(\frac{\pi}{2}-J,\frac{\pi}{2}-h\right){\frak R}^{\cal F}_-(h).
\end{eqnarray}
Here we exploited the fact ${\sf V}^{\cal F}(\pi/2-J,\pi/2-h)={\sf V}^{\cal F}(-J,-h){\sf P}$ for an even number of sites $N$. Therefore, we found the required duality operators ${\frak R}_\pm^{\cal F}$ that help us map the couplings $(h,J)$ across the second critical line $h+J=\pi/2$ as shown in Fig. \ref{fig:floquetphase}. Crucially, on this line, the operators ${\frak R}_\pm^{\cal F}$ become symmetries of the Floquet evolution ${\sf V}^{\cal F}(h,\pi/2-h)={\sf V}^{\cal F}(\pi/2-J,J)$. The algebra of these non-invertible duality-symmetries can be extracted as
\begin{eqnarray}
    &&{\frak R}_+^{\cal F}(J)^2=\frac{1}{2}(\mathbb{1}+{\sf P}){\sf T}{\sf V}^{\cal F}\left(\frac{\pi}{2}-J,J\right),\nonumber\\
    &&{\frak R}_-^{\cal F}(h)^2=\frac{1}{2}(\mathbb{1}+{\sf P}){\sf T}{\sf V}^{\cal F}\left(h,\frac{\pi}{2}-h\right)^{\dagger}.
\end{eqnarray}
Moreover, from \eqref{eq:GenKWAlg}, we have $\mathfrak{R}_\pm^{\cal F}(x)^\dagger{\frak R}_\pm^{\cal F}(x)=(\mathbb{1}+{\sf P})/2$. Hence, on the critical line $h+J=\pi/2$, the non-invertible duality-symmetries ${\frak R}_\pm^{\cal F}$ act as the appropriate half-translation operators along the light-cone coordinate $x\pm t$, with the time evolution generated by the critical Floquet unitary ${\sf V}^{\cal F}(h,\pi/2-h)={\sf V}^{\cal F}(\pi/2-J,J)$. It is now imperative to consider the limit of infinitesimally small couplings $h,J\ll 1$. {Note that, by construction}, although the unitary ${\sf V}^{\cal F}(\pi/2-J,\pi/2-h)\simeq e^{{\rm i}JH_A+{\rm i}hH_B}{\sf P}$ does not correspond to an infinitesimal continuous-time evolution on the full Hilbert space, it effectively does so when restricted to the even parity sector. As a result, the duality relations above and the associated algebra essentially reduce to the continuous-time counterparts \eqref{eq:KW} and \eqref{eq:KWalg}, as can be checked directly. 
\begin{figure}
    \centering
    \includegraphics[width=0.8\linewidth]{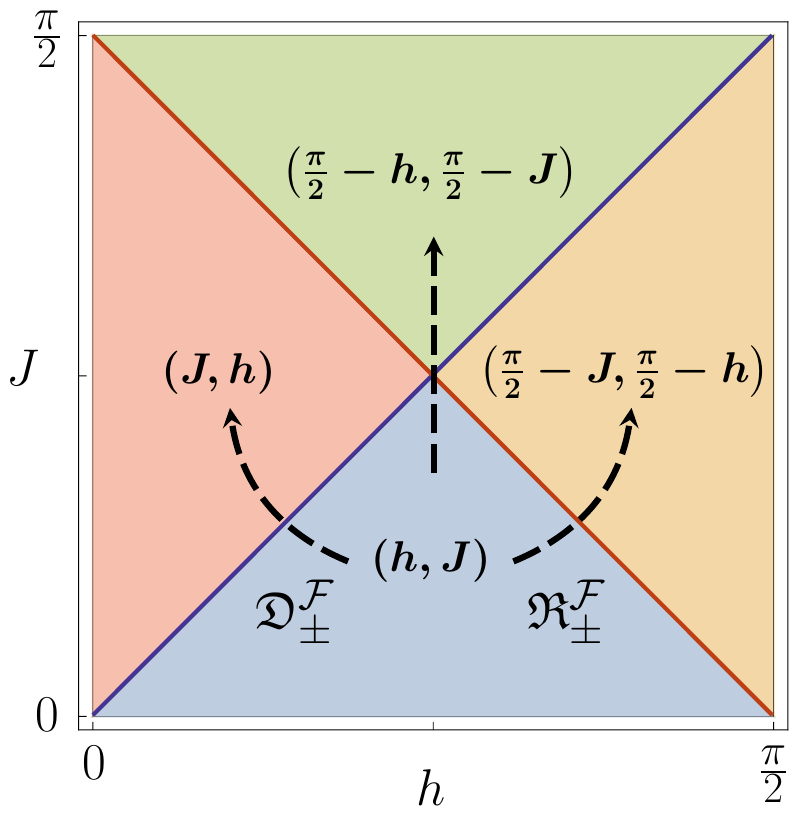}
    \caption{A schematic diagram explaining the actions of the non-invertible duality operators on the Floquet evolution \eqref{eq:Floquet2nd} in the interval $(h,J)\in[0,\pi/2]\times[0,\pi/2]$. As can be seen, the operators ${\frak D}^{\cal F}_\pm$ and ${\frak R}^{\cal F}_\pm$ relate distinct Floquet phases separated by the blue $(h=J)$ and the red $(h+J=\pi/2)$ critical lines, respectively. }
    \label{fig:floquetphase}
\end{figure}

We emphasize that one further can compose ${\frak R}_\pm^{\cal F}$ and ${\frak D}_\pm^{\cal F}$ as ${\frak R \frak D}^{\cal F}_\pm(x,y)={\frak R}_\pm^{\cal F}(x){\frak D}_\pm^{\cal F}(y)$, which implement
\begin{eqnarray}\label{eq:dual4}
    &&{\frak R\frak D}^{\cal F}_+(h,J){\sf V}^{\cal F}(h,J)={\sf V}^{\cal F}\left(\frac{\pi}{2}-h,\frac{\pi}{2}-J\right){\frak R\frak D}^{\cal F}_+(h,J),\nonumber\\
    &&{\frak R \frak D}^{\cal F}_-(J,h){\sf V}^{\cal F}(h,J)={\sf V}^{\cal F}\left(\frac{\pi}{2}-h,\frac{\pi}{2}-J\right){\frak R \frak D}^{\cal F}_-(J,h),\nonumber\\
\end{eqnarray}
and hence map the couplings $(h,J)$ to $(\pi/2-h,\pi/2-J)$. Importantly, the order of the operators in the composition can be reversed, with the desired maps recovered by appropriately adjusting the arguments.

\subsubsection{Floquet phase-structure}
As in the trotterized case, we can use the non-invertible duality operators to analyze the phase-structure of the system. It is clear that each of the non-invertible operators ${\frak D_\pm^{\cal F},\frak R_\pm^{\cal F}}$ annihilates any state with odd-parity. Consider now a parity-even eigenstate $|E_{(h,J)}^{\cal F}\rangle$ of the Floquet evolution ${\sf V}^{\cal F}(h,J)$ as
\begin{eqnarray}
    {\sf V}^{\cal F}(h,J)|E_{(h,J)}^{\cal F}\rangle=e^{-{\rm i}E_{(h,J)}^{\cal F}}|E_{(h,J)}^{\cal F}\rangle,
\end{eqnarray}
where $\{E_{(h,J)}^{\cal F}\}$ are the quasienergies. Owing to the relations \eqref{eq:dual1},\eqref{eq:dual2},\eqref{eq:dual4}, one can then show that ${\frak D}^{\cal F}_+(J)|E_{(h,J)}^{\cal F}\rangle$, ${\frak R}^{\cal F}_+(J)|E_{(h,J)}^{\cal F}\rangle$ and ${\frak R\frak D}^{\cal F}_+(h,J)|E_{(h,J)}^{\cal F}\rangle$ are the eigenstates of the Floquet evolutions ${\sf V}^{\cal F}(J,h)$, ${\sf V}^{\cal F}(\pi/2-J,\pi/2-h)$ and ${\sf V}^{\cal F}(\pi/2-h,\pi/2-J)$, respectively, with the same quasienergy $E^{\cal F}_{(h,J)}$. As one might expect, similar relations hold for the operators ${\frak D_-^{\cal F},\frak R_-^{\cal F},\frak{RD}_-^{\cal F}}$ as well. Notably, all of the above states have even parity. We now show that these Floquet eigenstates exhibit order-disorder dualities, induced by the above non-invertible duality operators.

Note that the duality operators ${\frak D}^{\cal F}_\pm$ act on the local terms as ${\frak D}^{\cal F}_-(x)Z_j=X_{j}X_{j+1}{\frak D}^{\cal F}_-(x)$, ${\frak D}^{\cal F}_+(x)X_{j}X_{j+1}=Z_{j+1}{\frak D}^{\cal F}_+(x)$. However, ${\frak D}^{\cal F}_\pm$ do not admit such simple actions on $Z_j$ and $X_{j}X_{j+1}$, respectively. Furthermore, the actions of ${\frak R}^{\cal F}_\pm(x)$ on the same local terms are given by
\begin{eqnarray}\label{eq:eq:Flodual2actions}
    &&{\frak R}^{\cal F}_-(x)Z_j=-X_{j}X_{j+1}{\frak R}^{\cal F}_-(x),\nonumber\\
    &&{\frak R}^{\cal F}_+(x)X_{j}X_{j+1}=-Z_{j+1}{\frak R}^{\cal F}_+(x).
\end{eqnarray}
Thus, apart from the sign differences, the non-invertible operators ${\frak D}_\pm^{\cal F}$ and ${\frak R}_\pm^{\cal F}$ act identically on the above local terms. As expected, no analogous simple actions of ${\frak R}_\pm^{\cal F}$ exist for the local operators $Z_j$ and $X_{j}X_{j+1}$, respectively. With these relations at hand, we now can show
\begin{widetext}
    \begin{eqnarray}\label{eq:FloqueOrdDisord}
        \langle Z_{j}\cdots Z_{k-1}\rangle^{\cal F}_{(h,J)}=(-1)^{k-j}\langle Z_{j+1}\cdots Z_k\rangle^{\cal F}_{\left(\frac{\pi}{2}-h,\frac{\pi}{2}-J\right)}=\langle X_{j}X_k\rangle^{\cal F}_{(J,h)}=(-1)^{k-j}\langle X_{j}X_{k}\rangle^{\cal F}_{\left(\frac{\pi}{2}-J,\frac{\pi}{2}-h\right)}
    \end{eqnarray}
\end{widetext}
where the correlations are evaluated in the parity-even eigenstates, related to one another by the non-invertible dualities as discussed earlier. Recall that, $\langle X_j X_k\rangle$ and $\langle Z_{j+1}\cdots Z_k\rangle$ describe the parity-invariant order and disorder correlations, respectively. Consequently, \eqref{eq:FloqueOrdDisord} represents the appropriate order-disorder duality relation for the Floquet evolution ${\sf V}^{\cal F}(h,J)$. In particular, for $(h,J)$ describing a disordered phase with asymptotically non-vanishing disorder correlation
\begin{eqnarray}
    \lim_{|k-j|\to\infty}\langle Z_{j}\cdots Z_{k-1}\rangle^{\cal F}_{(h,J)}\neq 0,
\end{eqnarray}
the couplings $(\pi/2-h,\pi/2-J)$ also display disordered behavior. On the other hand, the couplings $(J,h)$ and $(\pi/2-J,\pi/2-h)$ acquire non-zero long-range order as
\begin{eqnarray}
    \lim_{|k-j|\to\infty}\langle X_{j} X_k\rangle^{\cal F}_{(J,h)/(\pi/2-J,\pi/2-h)}\neq 0.
\end{eqnarray}
Indeed, this is in agreement with the results obtained in \cite{khemani2016phase,PhysRevB.93.245145,von2016phase,berdanier2018floquet}. The point $(h,J)=(\pi/4,\pi/4)$ corresponds to a multicritical point where the four phases meet.

\subsubsection{Dualities in presence of disorder}
However, to properly realize these phases in practice, one needs to introduce disorder in the coupling coefficients. We wish to conclude this section with a brief discussion on the non-invertible dualities for such a disordered Floquet evolution ${\sf V}^{\cal F}({\frak h,\frak J})=e^{-{\rm i}H_A({\frak h})}e^{-{\rm i}H_B({\frak J})}$, with $H_A({\frak h})=-\sum_{i=1}^Nh_iZ_i,~H_B({\frak J})=-\sum_{i=1}^NJ_iX_iX_{i+1}$. The couplings are often chosen from log-normal distributions $({\frak h,\frak J})$, with site-independent means $\overline{\ln(h_i)}=r_{\frak h}$, $\overline{\ln(J_i)}=r_{\frak J}$ and standard deviations $\delta\ln(h_i)=\Delta_{\frak h}$, $\delta\ln(J_i)=\Delta_{\frak J}$. We can now introduce one of the disordered KW duality operator  ${\frak D}^{\cal F}_{+}({\frak J})={\sf D}e^{-{\rm i}H_B({\frak J})}$. Unfortunately, in contrast to the homogeneous case, ${\frak D}_+^{\cal F}({\frak J})$ does not quite interchange the couplings $(h_i,J_i)$ in the evolution ${\sf V}^{\cal F}({\frak h,\frak J})$. Rather it maps $h_i\to J_{i-1},~J_i\to h_{i}$, as can be checked easily. Nonetheless, since the physical results follow from averaging over many disorder realizations, the non-invertible KW duality operator effectively interchanges the distributions ${\frak D}_+^{\cal F}({\frak J}){\sf V}^{\cal F}({\frak h,\frak J})={\sf V}^{\cal F}({\frak J,\frak h}){\frak D}^{\cal F}_+({\frak J})$, as long as the distributions $({\frak h,\frak J})$ are site-independent. 
One now can construct the other duality operators ${\frak D}_-^{\cal F}({\frak h})={\sf D}e^{{\rm i}H_{A}({\frak h})},{\frak R}_+^{\cal F}({\frak J})\simeq {\frak X}{\frak D}_+^{\cal F}({\frak J}){\frak X},{\frak R}_-^{\cal F}({\frak h})\simeq {\frak Z}{\frak D}_-^{\cal F}({\frak h}){\frak Z}$ and establish similar results.

}

\section{Conclusion}\label{sec:conclu}
In this work we have reported the integrable trotterization of the critical TFIM from the perspective of QISM. We started with an integrable Majorana fermionic model, which can be derived from a well-defined, non-local, fermionic solution of the YBE. Notably, when the time evolution is discretized, it still remains integrable. The local conserved quantities, given by the logarithmic derivatives of the transfer matrix, are quadratic in terms of the fermions. We then used the JW transformation to map the above fermionic model to the critical TFIM. The discrete time evolution operator thus obtained can be considered as a sequence of local quantum gates. The trotterized circuit \eqref{eq:Gates} has a particular ordering, namely ${\sf V}=V_AV_B$, which is not the same as the other obvious trotterization $\overline{\sf V}=V_BV_A\neq {\sf V}$. We point out that, the circuit $\overline{V}$ can be obtained by our method if we reverse the signs of the inhomogeneity parameters in the monodromy operator \eqref{eq:inhomMonoIsing}. Along with the other mutually commuting conserved quantities, we found the appropriate generalizations of the well-known KW duality-symmetry operator for the trotterized evolution. These operators generate a much larger algebra than their continuous-time counterpart, involving both the time evolution and the spatial translation operators. We argue that the discretization of both space and time leads to such enhanced algebra. Furthermore we observed that the KW duality-symmetry operators also can map between two distinct trotterization schemes of different order. This can be seen as an application of discrete time versions of the KW duality-symmetry operators constructed here. {We further discussed the implications of the non-invertible duality-symmetries on the phase structure of the discrete-time evolution. After establishing the order-disorder duality in the trotterized circuit, we moved to the general Floquet setting. Interestingly, the Floquet evolution exhibits a richer phase structure, with additional phases that do not appear in the undriven system. We systematically construct the relevant non-invertible duality-symmetries and show how they relate different regions of the phase diagram. }

{An immediate generalization of the present work would be to investigate the open boundary conditions, both in the quantum circuit and in the Hamiltonian limit. Such open integrable systems often host \textit{strong zero modes} (SZMs). Originally introduced by Fendley \textit{et al.} \cite{fendley2016strong,kemp2017long} in the context of continuous-time evolution, the idea of such SZMs recently has been extended to the setting of Floquet quantum circuits \cite{vernier2024strong}. It is therefore natural to ask whether our construction can admit the presence of such SZMs. The appropriate framework for studying integrable systems with open boundary conditions is provided by the Sklyanin construction \cite{sklyanin1988boundary}. The central object in this framework is the $K$-operator, which carries a single index and satisfies the so-called \textit{reflection equation}. This is in sharp contrast to the Yang–Baxter $R$-operator, whose structure involves two indices. However, our fermionic $R$-operator \eqref{eq:R-op} differs from the usual solutions of the Yang-baxter equation in several ways. Crucially, the fermionic $R$-operator is non-local and does not satisfy the far-commutativity relation. Therefore, a rigorous description of integrable, open-boundary Majorana systems using the Sklyanin construction may require a fundamental study on its own. }

Throughout this work, we considered the non-invertible KW duality-symmetry operators which act on a specific Hilbert space and commute with the relevant time evolution. However, when regarded as a topological defect, the KW duality introduces local modification in the Hamiltonian \cite{grimm2002spectrum,henkel1989ising,baake1989ising,shao2023s}. In recent years, different aspects of such duality twisted Hamiltonians, ranging from the behavior of the entanglement entropy \cite{Roy:2021jus,Roy2022,Rogerson:2022yim} to the Floquet dynamics \cite{Mitra:2023xdo,10.21468/SciPostPhys.16.3.075,Yan:2024yrw} and implementation on quantum computer \cite{Samanta:2023fvs}, have been investigated extensively. We suspect that such KW duality-twisted Hamiltonian for the critical TFIM can be obtained by introducing local impurity in the transfer matrix. Furthermore, it would be interesting to investigate the discrete-time evolution of such Hamiltonian. In particular, we want to study the fate of integrable trotterization for this kind of systems. Looking at the fermionic picture, insertion of the impurity in the transfer matrix is expected to yield twisted boundary conditions in the Majorana chain. Such twisted boundary conditions often host Majorana zero modes \cite{kawabata2017exact}. It will be interesting to see if our method captures the presence of such zero modes.

{To conclude our work, we recall that as the time-interval of the discrete-evolution approaches to zero, the discrete-time KW-duality operators ${\frak D}_\pm$ essentially reduce to the continuous-time KW-duality operator ${\sf D}$. However, for a non-zero time-step their actions clearly differ. As a consequence, the corresponding topological defects are expected to have non-trivial actions. It is instructive to explore if twisting  ${\sf V}(h,J)$ by the defects ${\frak D}_\pm$ results in the same discrete-time evolution, obtained from the continuous-time evolution of the ${\sf D}$ defect Hamiltonian. We plan to pursue this direction in an immediate future work.}

\section*{Acknowledgements}
VK is funded by the U.S. Department of Energy, Office of Science, National Quantum Information Science Research Centers, Co-Design Center for Quantum Advantage ($C^2QA$) under Contract No. DE-SC0012704.

\bibliographystyle{ieeetr}
\bibliography{refs}

\appendix
\setcounter{equation}{0}
\onecolumngrid

\section{Commuting transfer matrices}\label{app:transfercommutation}
Here we give the details on the commutativity of the transfer matrices, which follows from \eqref{eq:RTT}. Since we need to take partial trace over the auxiliary indices, it is desirable to look for a factorized representation of the Majorana modes. We begin with distinguishing the auxiliary Majorana modes ${\cal G}_{ab}=\{\gamma_a,\gamma_b\}$ from the physical Majorana modes ${\cal G}_{\rm P}=\{\gamma_j|j=1,\cdots,{2N}\}$. Furthermore, assume ${\cal G}_{ab}$ and ${\cal G}_{\rm P}$ have representations $\Phi_{ab}$ and $\Phi_{\rm P}$ on the Hilbert spaces ${\cal H}_{ab}$ and ${\cal H}_{\rm P}$, respectively
\begin{eqnarray}
    &\{\Phi_{ab}(\gamma_a),\Phi_{ab}(\gamma_b)\}=2\delta_{a,b}{\mathbb 1}_{{ab}},&\nonumber\\
    &\quad \{\Phi_{\rm P}(\gamma_j),\Phi_{\rm P}(\gamma_k)\}=2\delta  _{jk}{\mathbb 1}_{\rm P},\quad j,k=1,2,\cdots,{2N}.&
\end{eqnarray}
where ${\mathbb 1}_{ab}$ and ${\mathbb 1}_{\rm P}$ are the identity operators on the respective Hilbert spaces. Then one possible representation of the algebra generated by ${\cal G}={\cal G}_{ab}\cup{\cal G}_{\rm P}$ is given by
\begin{eqnarray}
&&\Phi(\gamma_{a(b)})=\Phi_{ab}(\gamma_{a(b)})\otimes\mathbb{1}_{\rm P},\nonumber\\
&&\Phi(\gamma_j)={\rm i}\Phi_{ab}(\gamma_a\gamma_b)\otimes\Phi_{\rm P}(\gamma_j).
\end{eqnarray}
This is a valid representation on ${\cal H}={\cal H}_{ab}\otimes{\cal H}_{\rm P}$, as can be checked easily. It satisfies the required algebra
\begin{eqnarray}
    \{\Phi(\gamma_\mu),\Phi(\gamma_\nu)\}=2\delta  _{\mu\nu}{\mathbb 1},\quad\mu,\nu=a,b,1,2,\cdots,{2N}.
\end{eqnarray}
The important property of the above representation is that the auxiliary Majorana modes act trivially on the Hilbert space ${\cal H}_{\rm P}$. We therefore consider ${\cal H}_{\rm P}$ as the relevant physical Hilbert space. However, the physical Majorana modes, $\gamma_j$'s act non-trivially on the auxiliary Hilbert space ${\cal H}_{ab}$, capturing the non-local nature of the Majorana modes. Before going further, we note that, the representations $\Phi_{ab}(\gamma_a),\Phi_{ab}(\gamma_a)$ and $\Phi_{ab}(\gamma_a\gamma_b)$ are traceless. This directly follows from the anticommutation relation and the cyclicity of the trace. To see this, let us consider
\begin{eqnarray}
    {\rm tr}_{{\cal H}_{ab}}[\Phi_{ab}(\gamma_a)]&=&{\rm tr}_{{\cal H}_{ab}}[\Phi_{ab}(\gamma_a)\Phi_{ab}(\gamma_b)\Phi_{ab}(\gamma_b)]\quad\quad~~\,\text{(as $\gamma_b^2=1$)}\nonumber\\
    &=&-{\rm tr}_{{\cal H}_{ab}}[\Phi_{ab}(\gamma_b)\Phi_{ab}(\gamma_a)\Phi_{ab}(\gamma_b)]\quad\quad\text{(due to anticommutation)}\nonumber\\
    &=&-{\rm tr}_{{\cal H}_{ab}}[\Phi_{ab}(\gamma_b)\Phi_{ab}(\gamma_b)\Phi_{ab}(\gamma_a)]\quad\quad\text{(due to cyclicity of trace)}\nonumber\\
    &=&-{\rm tr}_{{\cal H}_{ab}}[\Phi_{ab}(\gamma_a)].
\end{eqnarray}
This immediately implies ${\rm tr}_{{\cal H}_{ab}}[\Phi_{ab}(\gamma_a)]=0$. Similarly, ${\rm tr}_{{\cal H}_{ab}}[\Phi_{ab}(\gamma_b)]=0={\rm tr}_{{\cal H}_{ab}}[\Phi_{ab}(\gamma_a\gamma_b)]$.

Now one can multiply both the sides of the $RTT$-relation \eqref{eq:RTT} by $R_{a,b}(\lambda-\mu)^{-1}$ from either left or right and then perform the partial trace over ${\cal H}_{ab}$. Since $\Phi(R_{a,b}(\lambda))$ acts trivially on ${\cal H}_{\rm P}$, we can use the cyclicity property of the trace to obtain
\begin{eqnarray}\label{eq:preComTrans}
    {\rm tr}_{{\cal H}_{ab}}\Phi\left(T_a(\lambda|\{\eta\})T_b(\mu|\{\eta\})\right)={\rm tr}_{{\cal H}_{ab}}\Phi\left(T_b(\mu|\{\eta\})T_a(\lambda|\{\eta\})\right).
\end{eqnarray}
Let us expand the monodromy operators as
\begin{eqnarray}
    T_{a(b)}(\lambda|\{\eta\})=\tau(\lambda|\{\eta\})+\gamma_{a(b)}\sigma(\lambda|\{\eta\}).
\end{eqnarray}
Owing to the even number of Majorana modes and the structure of the  monodromy operator \eqref{eq:inhomMono}, the operators $\tau(\lambda|\{\eta\})$ and $\sigma(\lambda|\{\eta\})$ contain even and odd powers of the physical Majorana modes $\gamma_j,~j=1,\cdots,2N$, respectively. Consequently, the representation takes the form
\begin{eqnarray}
    &&\Phi\left({\tau}(\lambda|\{\eta\})\right)={\mathbb 1}_{{ab}}\otimes\Phi_{\rm P}\left(\tau(\lambda|\{\eta\})\right),\quad \Phi\left({\sigma}(\lambda|\{\eta\})\right)\,={\rm i}\Phi_{ab}(\gamma_a\gamma_b)\otimes\Phi_{\rm P}\left(\sigma(\lambda|\{\eta\})\right).
\end{eqnarray}
This leads, from \eqref{eq:preComTrans}, to the commuting transfer matrices as 
\begin{eqnarray}
    [\Phi_{\rm P}\left(\tau(\lambda|\{\eta\})\right),\Phi_{\rm P}\left(\tau(\mu|\{\eta\})\right)]=0,\quad \Phi_{\rm P}\left(\tau(\lambda|\{\eta\})\right)={\rm tr}_{{\cal H}_{ab}}\Phi\left(T_{a(b)}(\lambda|\{\eta\})\right).
\end{eqnarray}
However, often we shall drop the labels of the representations and write down simply
\begin{eqnarray}
    [\tau(\lambda|\{\eta\}),\tau(\mu|\{\eta\})]=0,\quad\tau(\lambda|\{\eta\})={\rm tr}_{{\cal H}_{ab}}\left[T_{a(b)}(\lambda|\{\eta\})\right].
\end{eqnarray}
This completes the proof of the commutativity of the transfer matrices.

\section{Local conserved quantities}\label{app:LocCon}
Here we list down some of the local conserved quantities for the trotterized quantum circuit \eqref{eq:intcirc}. First we recall the local conserved quantities for the completely homogeneous case, namely which commute with the Hamiltonian \eqref{eq:HamMajInt}. These charges can be obtained by taking the higher order logarithmic derivatives of the completely homogeneous transfer matrix as
\begin{eqnarray}
    &&Q_r=\frac{{\rm d}^r}{{\rm d}\lambda^r}\ln\tau(\lambda,0)\Big|_{\lambda=0}={\rm i}\sum_{j=1}^{2N-r}\Gamma_j\Gamma_{j+r}+{\cal B}_r,\quad r\geq 2,
\end{eqnarray}
where ${\cal B}_r=-{\rm i}\sum_{k=1}^r\Gamma_{2N-r+k}\Gamma_{2N+k}$ is the boundary term, having support over $2r$-sites around the boundary, namely over the indices $2N-r+1$ to $2N+r$. These charges can also be obtained by manipulating the infinite-lattice charges derived in \cite{Sinha:2025wqf} and then by adding suitable boundary terms to preserve the twisted translation symmetry. 

Now the conserved charges for the inhomogeneous case are obtained by taking the logarithmic derivative of the inhomogeneous transfer matrix as \cite{vanicat2018integrable}
\begin{eqnarray}
   Q^{(r)}_\pm(\omega)=\frac{{\rm d}^r}{{\rm d}\lambda^r}\ln \tau\left(\lambda,\omega\right)\Big|_{\lambda=\pm\frac{\omega}{2}}.
\end{eqnarray}
Carrying out the derivative, we find the explicit expressions (up to some constant factors depending on $\omega$) of $Q^1_\pm(\omega)$ as
\begin{eqnarray}
    &&Q^{(1)}_\pm(\omega)={\rm sech}^2(\omega)Q_1\mp2\tanh(\omega)M^{(1)}_\pm,\nonumber\\
    &&M^{(1)}_+={\rm i}\sum_{j=1}^{N-1}\Gamma_{2j}\Gamma_{2j+2}-{\rm i}~\Gamma_{2N}\Gamma_2,\nonumber\\
    &&M^{(1)}_-={\rm i}\sum_{j=1}^{N-1}\Gamma_{2j-1}\Gamma_{2j+1}-{\rm i}~\Gamma_{2N-1}\Gamma_1.
\end{eqnarray}
When the inhomogeneity goes to zero, we recover our Hamiltonian from these local charges $Q^{(1)}_\pm(0)=H$. The next charge $Q^2_\pm(\omega)$ can be obtained as
\begin{eqnarray}
    &&Q_\pm^{(2)}(\omega)=\pm{\rm sech}(2\omega)\tanh(2\omega)\left(Q_1-Q_3\right)+{\rm sech}(2\omega)^2Q_2+\tanh(2\omega)^2M^{(2)}_\pm,\nonumber\\
    &&M^{(2)}_+={\rm i}\sum_{j=1}^{N-2}\Gamma_{2j}\Gamma_{2j+4}-{\rm i}\Gamma_{2N-2}\Gamma_2-{\rm i}\Gamma_{2N}\Gamma_4,\nonumber\\
    &&M^{(2)}_-={\rm i}\sum_{j=1}^{N-2}\Gamma_{2j-1}\Gamma_{2j+3}-{\rm i}\Gamma_{2N-3}\Gamma_1-{\rm i}\Gamma_{2N-1}\Gamma_3.
\end{eqnarray}
Evidently, we have $Q^{(2)}_\pm(0)=Q_2$, as expected. Notably, all the charges are quadratic in nature. We expect this to be true for $r\geq 3$ also.


\section{Onsager algebra}\label{app:Onsager}
The Onsager algebra \cite{onsager1944crystal} is an infinite-dimensional Lie algebra, spanned by $\{A_m,G_m\},m\in\mathbb{Z}$, satisfying
\begin{eqnarray}
    &[A_l,A_m]=4G_{l-m},\quad[G_l,G_m]=0,&\nonumber\\
    &[G_l,A_m]=2A_{m+l}-2A_{m-l}.&
\end{eqnarray}
In the context of the TFIM, the first two charges are given by
\begin{eqnarray}
    A_0=\sum_j Z_j,\quad A_1=\sum_jX_jX_{j+1}.
\end{eqnarray}
They satisfy the celebrated Dolan-Grady condition \cite{dolan1982conserved}, which essentially is a recursive structure among the commutators between $A_0$ and $A_1$, reading as 
\begin{eqnarray}\label{eq:D-G condition}
    &[A_0,[A_0,[A_0,A_1]]]=16[A_0,A_1],&\nonumber\\
    &[A_1,[A_1,[A_1,A_0]]]=16[A_1,A_0].&
\end{eqnarray}
As a result, the above two charges are sufficient to generate the Onsager algebra \cite{perk1989star,davies1990onsager}. Exploiting the above algebra, it is possible to construct a set of mutually commuting conserved charges. In particular, the Hamiltonian $H_x=A_0+{\cal J}A_1$ belongs to the family of the commuting charges
\begin{eqnarray}
    Q^{(m)}_{\cal J}=A_m+A_{-m}+{\cal J}(A_{1+m}+A_{1-m}),\quad H_{\cal J}=Q^{(0)}_{\cal J}.
\end{eqnarray}
The critical TFIM is recovered at ${\cal J}=1$. Then the question naturally arises as to whether we can deduce the charges $A_0$ and $A_1$ using our mechanism. However, since $[A_0,A_1]\neq 0$, it is clear that they cannot be obtained from one specific transfer matrix. Interestingly, allowing the presence of a second transfer matrix turns out to be sufficient to accomplish the task. 

To see this, consider two different inhomogeneous transfer matrices $\tau(\lambda|\pm\omega)$. Note that $[\tau(\lambda|\omega),\tau(\mu|-\omega)]\neq 0$ in general. Furthermore, with the identification $\tan(\beta)=\tanh(\omega)$, we have
\begin{eqnarray}
    &&\tau\left(\frac{\omega}{2}\Big|-\omega\right)^{-1}\tau\left(-\frac{\omega}{2}\Big|\omega\right)=e^{2{\rm i}N\beta}e^{-2\beta\sum_{j}\Gamma_{2j-1}\Gamma_{2j}},\nonumber\\
    &&\tau\left(-\frac{\omega}{2}\Big|-\omega\right)^{-1}\tau\left(\frac{\omega}{2}\Big|\omega\right)=e^{-2{\rm i}N\beta}e^{2\beta\sum_{j}\Gamma_{2j}\Gamma_{2j+1}}.
\end{eqnarray}
If we now represent the Majorana fermions in terms of the spin-1/2 operators using the JW transformation, the charges $A_0$ and $A_1$ can be extracted, apart from a boundary term, from the above relations as
\begin{eqnarray}\label{eq:OnsagerA0A1}
   && A_0=N{\mathbb 1}+{\rm i}\frac{\log\left[\tau\left({\omega}/{2}|-\omega\right)^{-1}\tau\left(-{\omega}/{2}|\omega\right)\right]}{2\arctan{(\tanh(\omega))}},\nonumber\\
    &&A_1=N{\mathbb 1}-{\rm i}\frac{\log\left[\tau\left(-{\omega}/{2}|-\omega\right)^{-1}\tau\left({\omega}/{2}|\omega\right)\right]}{2\arctan{(\tanh(\omega))}}.
\end{eqnarray}
Since $A_0$ and $A_1$ are sufficient to obtain the complete Onsager algebra, the relations in \eqref{eq:OnsagerA0A1} essentially provides the necessary connection between the transfer matrix formalism and the Onsager algebra.

\end{document}